\documentclass[a4paper,11pt]{article}
\pdfoutput=1
\usepackage{jheppub}
\usepackage{graphicx}
\usepackage{amsmath}
\usepackage{amsfonts}
\usepackage{amssymb}
\usepackage{slashed}
\usepackage[colorlinks=true,linkcolor=blue]{hyperref}
\usepackage{bm}
\usepackage{empheq}
\usepackage[super]{nth}
\usepackage{siunitx}
\usepackage{subcaption}

\def\po{\phantom{1}}

\def\ltotg{\ell_{\mathrm{tot},\gamma}}
\def\ltot{\ell_{\mathrm{tot}}}

\def\LL{\mathcal{L}}
\def\be{\begin{equation}}
\def\ee{\end{equation}}
\def\bea{\begin{eqnarray}}
\def\eea{\end{eqnarray}}

\def\Eq#1{Eq.~\eqref{#1}}

\title{Global cosmic string networks as a function of tension}
\author{Vincent B.\ Klaer, Guy D.\ Moore}

\affiliation{Institut f\"ur Kernphysik, Technische Universit\"at Darmstadt\\
Schlossgartenstra{\ss}e 2, D-64289 Darmstadt, Germany}
\emailAdd{vklaer@theorie.ikp.physik.tu-darmstadt.de,guy.moore@physik.tu-darmstadt.de}

\abstract{
We investigate the properties of global cosmic string networks
as a function of the ratio of string tension to Goldstone-field
coupling, and as a function of the Hubble damping strength.
Our results show unambiguously that the string density is sensitive to
this ratio.  We also find that existing semi-analytical
(one-scale) models must be missing some important aspect of the
network dynamics.  Our results point the way towards improving such
models.
}

\begin{document}
\maketitle
\section{Introduction}
\label{sec:intro}

Cosmic strings are an interesting consequence of any
beyond-Standard-Model theory in which a U(1) symmetry spontaneously
breaks.  They could play a role in modern cosmology -- modern
constraints show that they play a negligible role in cosmic structure
formation
\cite{Ade:2013xla,Urrestilla:2011gr,Lizarraga:2014xza,Lazanu:2014xxa,%
Lopez-Eiguren:2017dmc},
but they could contribute to the
gravitational radiation background
\cite{Vachaspati:1984gt,Blanco-Pillado:2017rnf}.
Perhaps most
interesting, axionic cosmic strings may play a pivotal role in
establishing the density of axions in the Universe
\cite{Davis:1986xc,Harari:1987ht,Hagmann:1998me,Battye:1993jv,Battye:1994au,%
  Yamaguchi:1998gx,Yamaguchi:1999yp,Hiramatsu:2010yu,%
  Hiramatsu:2012gg,axion1,axion4},
and therefore the relation between the axion dark matter abundance and
the axion mass \cite{Visinelli:2014twa}.
And they determine the level of short-scale inhomogeneities in the
axion dark-matter background, which could lead to small, very
overdense axionic dark matter features (axion miniclusters or
axion stars) \cite{Hogan:1988mp,Kolb:1993zz,Kolb:1993hw,Zurek:2006sy,%
Hardy:2016mns,Enander:2017ogx,Levkov:2018kau,Vaquero:2018tib}.

Most research to date has focused on local cosmic strings, meaning
strings where the spontaneously broken U(1) symmetry couples to a U(1)
gauge field \cite{Kibble:1976sj} or otherwise leads to no
massless degrees of freedom, and therefore no long-range interactions
between strings.  Large-scale numerical
simulations
\cite{Bennett:1989yp,Allen:1990tv,%
Vanchurin:2005yb,Olum:2006ix,BlancoPillado:2011dq,Hindmarsh:2017qff}
generally treat
this case, and analytical models designed to describe string networks
\cite{Martins:1996jp,Martins:2000cs,Martins:2003vd,Martins:2018dqg}
are generally fitted to these simulations.  It appears
that such analytical models now do quite a good job in describing the
core features of these string networks.

But axions would arise from global cosmic string networks, meaning
strings arising from a U(1) symmetry breaking where no gauge bosons
receive a mass \cite{Hindmarsh:1994re,vilenkin2000}.  Instead, these networks
couple to massless Goldstone bosons, which can be radiated from the
network and which can communicate long-range interactions between
strings.  Effective theories governing these networks are known
\cite{Dabholkar:1989ju}, but existing large-scale numerical simulations
\cite{Hiramatsu:2010yu,axion1,Vaquero:2018tib,Buschmann:2019icd}
face resolution issues, which limit
their reliability.  Specifically, the string evolution is a two-scale
problem:  there is an infrared scale, set by the light-crossing
distance of the spacetime, which is of order the Hubble scale $H^{-1}$.
The inter-string separation is of order this scale.
And there is a microphysical scale, of order
the inverse mass of the heavy fields which must exist in the model
giving rise to the string.  For instance, if the string arises from a
microphysical scalar theory with a Mexican-hat potential, this is the
inverse of the mass scale for radial excitations $m$.  This scale sets
the size of the string's core, $r_{\mathrm{core}} \sim 1/m$.
The string's gradient energy density scales with distance from the
string core as $1/r^2$, and therefore the string tension is
logarithmically dependent on this scale hierarchy:  the tension
$T \simeq \pi f^2 \int_{1/m}^{1/H} r\,dr/r^2 \simeq \pi f^2 \ln(m/H)$,
with $f$ the symmetry-breaking scale and $\ln(m/H) \equiv \kappa$ the
logarithm of this scale ratio.  Note however that the strength of the
string's interactions with long-range Goldstone boson modes is
proportional to $f^2$ but without this factor of $\kappa$.  Therefore
$f$ cancels out in establishing the network dynamics, but $\kappa$
does not, and the network dynamics could show logarithmic sensitivity
to this ratio of scales.

Some numerical results observe such a sensitivity
\cite{axion1,Gorghetto:2018myk,Kawasaki:2018bzv,%
  Vaquero:2018tib,Buschmann:2019icd}, while others, including recent results,
do not \cite{Hindmarsh:2019csc}.  This discrepancy needs to be resolved.
We also need simulations which achieve much larger scale hierarchies
than straightforward scalar-field simulations can reach.
For instance, for axions at the QCD epoch we have
$m \sim f_a \sim 10^{11}\:\mathrm{GeV}$ and
$H \sim T^2/m_{\mathrm{pl}}
\sim (0.3\:\mathrm{GeV})^2/10^{19}\:\mathrm{GeV}$, and therefore
$\kappa \sim 70$; whereas in numerical studies $m$ must be smaller
than the inverse lattice spacing and $H$ must be larger than the
inverse box size, constraining (roughly) $\kappa \leq 7$.
Multigrid methods \cite{Drew:2019mzc} might be able to reach
spacing-to-separation ratios of order $10^6$, giving $\kappa \sim 14$,
but this is still severely insufficient.

We see a need to improve this situation with better simulations of
global string networks.  These would allow us to confront analytical
models with data to see where they come up short.  Recently, a similar
effort aimed at improving simulations of domain wall networks showed
significant shortcomings in the existing analytical models
\cite{Martins:2016wqq}.

In this work we will address this problem with two sorts of
simulations.  First, we will perform simulations where the ratio of
the Hubble scale to the string core size is held fixed.  These are
unphysical but they have the virtue that a true scaling solution
exists, and lattice-spacing artifacts and initial-condition transients
rapidly disappear.  Using these simulations, we will demonstrate
robustly, within scalar-only simulations, that the string network
really does depend significantly on $\kappa$.

Next, we will use our recently introduced numerical method for
introducing a large string tension into global string networks, to
see how network properties depend on tension at large tension values.
These studies rapidly run into resolution problems \textsl{if} we
stick with the radiation-dominated FRW metric.  But 
by considering matter-dominated FRW metrics and metrics for universes
with a slightly negative-pressure equation of state, we can explore
networks with much stronger Hubble drag.  This Hubble drag gives an
additional variable for comparison with analytical models.  It also
destroys kinks and other small-scale structures on strings, which
makes it easier to achieve scaling in the string network
evolution.  Furthermore, as we will discuss, existing one-scale models
\cite{Martins:1996jp,Martins:2000cs} predict that the behavior of
global and local networks will rapidly become more similar as the
Hubble drag is increased -- a prediction we will test.

The next section sets the stage by reviewing our expectations for
global string networks, based on analytical ``one-scale'' models.
Then Section \ref{sec:fixkappa} will explore networks in the radiation
era with a small
but fixed $\kappa$ value, as a function of $\kappa$.  We show robustly
that the network density is sensitive to this measure of the scale
hierarchy.  Then we present our methodology and results for studying
higher tension networks with varying Hubble drag, looking at
the string velocity and the kinkiness of strings as well as the
network density.  We also present results for the rate of string loop
production.  Section
\ref{sec:discussion} discusses what we have learned, and in
particular, the implications of these results for existing analytical
models.  Our most notable findings are that global string networks have
\textsl{larger} average string velocity than local networks, opposite
to the predictions of the one-scale model
\cite{Martins:1996jp,Martins:2000cs}, and that the
string density does \textsl{not} rapidly approach that of a local
network as one increases the Hubble drag.  We believe that some work
is needed on the modeling side, probably by incorporating the effects
of long-range inter-string interactions.

\section{Analytical expectations}
\label{sec:onescalemodel}

This section reviews the analytical one-scale model of string
networks, originally introduced by Martins and Shellard
\cite{Martins:1996jp} and later extended for global string networks
\cite{Martins:2000cs}.

First we quickly review the cosmological background spacetimes we will
consider.  We assume an FRW universe with equation of
state $P = w \varepsilon$.  The physically most interesting cases are
$w=0$ (matter domination) and $w=1/3$ (radiation domination), but we
will allow all values $w \in [-1/3,1]$.  Values $w>1$ violate
the dominant energy condition and do not appear to arise from any
state of a renormalizable field theory.  Values $w<-1/3$ do not allow
for string network scaling solutions.  But by considering values
$w \in [-1/3,0]$, we will be able to examine string networks with
particularly tractable dynamics, which challenge one-scale models.

The two Einstein equations describing the evolution of the energy
density $\varepsilon$ and scale factor $a$ as a function of $t$ for
such a metric are
\begin{equation}
  \label{Friedmann}
  \frac{1}{\varepsilon} \frac{d\varepsilon}{dt} = -3(1+w) H \,,
  \qquad
  H \equiv \frac{da}{a dt} = \frac{\sqrt{\varepsilon}}{M_{\mathrm{pl}}}
\end{equation}
with $M_{\mathrm{pl}}$ the reduced Planck mass.  We also introduce the
conformal time
\begin{equation}
  \label{conformaltime}
  d\tau = \frac{1}{a} dt
\end{equation}
in terms of which the metric is
$g_{\mu\nu} = a^2(\tau) \eta_{\mu\nu}$.
After some straightforward algebra we find
\begin{equation}
  \label{Handt}
  \frac{a}{a_0} = \left( \frac{t}{t_0} \right)^{\frac{2}{3(1+w)}} \,,
  \qquad
  H = \frac{2}{3(1+w)} t^{-1} \,, \quad \mbox{and} \quad
  \tau \propto a^{\frac{1+3w}{2}} .
\end{equation}
For future reference, we can work in comoving coordinates and
conformal time, in which case the metric scales with conformal time as
\begin{equation}
  \label{g-in-tau}
  g_{\mu\nu} = a^2(\tau) \eta_{\mu\nu} = \tau^n \eta_{\mu\nu} \,,
  \qquad n \equiv \frac{4}{1+3w} \,.
\end{equation}
We will see that $n$ parameterizes how strongly Hubble drag acts to
slow down cosmic strings.%
\footnote{For comparison with other literature: some
authors \cite{Correia:2019bdl} work in terms of $m \equiv Ht$,
which is related to $n$ via $m=n/(2+n)$ or $n=2m/(1-m)$.}
By considering values of $w$ from 1 to $-1/3$ we can arrange for $n$ to
take values between 1 and $\infty$, though values $n>4$ require values
$w<0$ which are difficult to achieve with physically sensible
cosmological fluids.

We now return to the issue of describing a string network.  One-scale
models \cite{Martins:1996jp,Martins:2000cs} postulate that a string
network is described by the mean inter-string separation $L$ and the
mean string velocity $v$.  We should understand $L$ in terms of the
string length per unit volume, $\ltotg/V = 1/L^2$; here
$\ltotg$ is the total invariant length of string in a
large volume $V$, and invariant length means that we weight a string's
length with a factor of $\gamma \equiv 1/\sqrt{1-v^2}$ so that string
energy, rather than geometrical string length, is the relevant
quantity; $\ltotg = \int \gamma d\sigma$ where the integral is over all
string and $d\sigma$ is the differential geometrical length of string.
The velocity $v$ should be understood as
the RMS string velocity of the network, weighted by
string energy and excluding small loops.  Because strings are curved,
string tension tends to accelerate strings and increase $v$ provided
that $v^2 < 1/2$; this effect gets stronger the smaller $L$ is.
Hubble drag slows down the network in a way which depends only on the
value of $H$.  Martins and Shellard argue for velocity evolution
obeying
\begin{equation}
  \label{v-evolution}
  \frac{dv}{dt} = (1-v^2) \left( \frac{k(v)}{L}
  - 2H v \right) \,,
\end{equation}
where $k(v)$ is a velocity-dependent coefficient reflecting how much
the string curvature accelerates the string motion.
Analytic estimates give $k=2\sqrt{2} / \pi$, but it is better to treat
it as a velocity-dependent function to be fit to network evolution data.
Recently Correia and Martins \cite{Correia:2019bdl} have extended the
model by considering $k(v)$ to be a more general function of velocity.
Their fit to the abelian Higgs model at a range of $n$ values
from $n=2$ to $n=38$ obtains
\begin{equation}
  \label{correia1}
  k(v) = k_0 \frac{1-(qv^2)^\beta}{1+(qv^2)^\beta} \,,
  \qquad
  k_0 \simeq 1.37, \;
  q \simeq 2.3, \;
  \beta \simeq 1.5 \,.
  \end{equation}
We will use this form in what follows.

For a true scaling solution, we expect $dv/dt = 0$, in which case
\Eq{v-evolution} reduces to $L = k(v)/2Hv$.  As $k(v)/v$ is a strictly
decreasing function of $v$, this predicts that smaller velocities
correspond with larger $L$ values and therefore lower-density
networks.  It is also useful to introduce the conformal correlation
length $\xi \equiv L/a$; in terms of conformal length and time we have
\begin{equation}
  \label{dv_conformal}
  a\tau = \frac{n+2}{2} t \,, \quad
  Ht = \frac{n}{n+2} \quad \Rightarrow \quad
  \frac{L}{a\tau} = \frac{\xi}{\tau} = \frac{1}{n} \: \frac{k(v)}{v} \,.
\end{equation}
This version is useful for identifying the network density in
conformal time, which is typically used in lattice simulations.
To convert to regular time, use $\xi/\tau = (L/t)\times 2/(n+2)$.

The other network evolution equation tells how fast the amount of
string changes.  Note that $\ltotg \propto 1/L^2$, and therefore
$(1/\ltotg) d\ltotg/dt = -(2/L) dL/dt$; so something which reduces the
amount of string increases $L$.  Therefore the rate at which $L$
changes is
\cite{Martins:1996jp,Martins:2000cs}
\begin{equation}
  \label{L-evolution}
  2 \frac{dL}{dt} = 2HL ( 1 + v^2 ) + F(v) \,, \qquad
  F(v) = c_{\mathrm{loop}} v   + d (k_0 - k(v))^r
  + \frac{s_{\mathrm{global}} v^6}{\kappa} \,.
\end{equation}
Here $2HL(1+v^2)$ is the loss of string density due to expansion, with $1$
arising from dilution of the string network and $v^2$ from loss of
string energy due to Hubble drag.  The term $F(v)$ represents all
mechanisms for string energy to convert into some other type of
energy, such as the production of string loops
$c_\mathrm{loop} v$, radiation of heavy modes represented
phenomenologically as $d (k_0-k(v))^r$, and radiation of Goldstone
modes $s_{\mathrm{global}} v^6/\kappa$.
We will again follow \cite{Correia:2019bdl} and choose
\begin{equation}
  \label{Correia2}
  c_{\mathrm{loop}} = 0.34, \quad
  d = 0.22, \quad
  r = 1.8, \quad
  s_{\mathrm{global}} = 70 \,.
\end{equation}
The value $s_{\mathrm{global}}=70$ is from
Ref.~\cite{Martins:2018dqg}, since this parameter was not considered
in Ref.~\cite{Correia:2019bdl}.  The inverse dependence on $\kappa$
arises because energy loss through Goldstone boson radiation is
proportional to $f^2$, but the string tension is proportional to
$f^2 \kappa$; therefore, the larger the $\kappa$ value is, the less
importance Goldstone radiation has to the string network's energy
budget.  The $v^6$ behavior is a generic prediction for emission of
massless, relativistic waves from a nonrelativistic source
\cite{Martins:2000cs}.
It is also conceivable that
some coefficients, particularly $c_{\mathrm{loop}}$, are different for
global than local networks; and $c_{\mathrm{loop}}$ may also have
velocity dependence which is not
reflected above.

Returning to \Eq{L-evolution}, the scaling expectation is
$L \propto t$, and therefore $dL/dt = L/t$ would
represent a scaling solution.  Again using $Ht = n/(n+2)$,
rearranging, and converting to conformal units, we find
\begin{equation}
  \label{Levolve2}
  (2-nv^2) \frac{\xi}{\tau} = F(v) \qquad \mbox{or} \qquad
  \frac{\xi}{\tau} = \frac{F(v)}{2-nv^2} \,.
\end{equation}
We can determine the scaling solutions for $v$ and $\xi/\tau$
by plotting \Eq{dv_conformal} and \Eq{Levolve2} in the $v,\xi/\tau$
plane and finding the intersection point.  But we can already see that
the presence of
$s_{\mathrm{global}}$ increases the value of $L/t$ and therefore
decreases $v$.  Therefore, two robust predictions of the one-scale
model are that global networks have smaller values of $v^2$ and larger
values of $L/t$ (lower network densities).
And for $n>4$ the prefactor on $\xi/\tau$ in \Eq{Levolve2}
vanishes for $v^2 = 2/n$, which is therefore an upper bound on
the string mean-squared velocity.  For values of $w \simeq -1/3$ the
network velocity becomes small, which makes the friction term
$\propto v^6$ irrelevant.  Therefore the one-scale model as formulated
above predicts that global and local networks should behave almost
identically in this regime.

To test this one-scale model, we should consider \textsl{both}
different values of $\kappa$ \textsl{and} different values of $w$ or
equivalently $n$, including $0 > w > -1/3$ or $n>4$.  Another
advantage of considering $w<0$ is that, in this regime,
strong Hubble drag tends to round off kinks and damp cusps, making
strings smoother and more easily resolved by lattice studies.
In preparation for numerical results on these quantities, we show in
Table \ref{table:onescale} what the one-scale model predictions are
for local and global string network densities
$\zeta \equiv (\xi/\tau)^{-2}$ and mean-squared velocities $v^2$ at
a few $\kappa$ values and for the three $n$ values which we will
consider further in later sections.  Note that we express the network
density in comoving coordinates and conformal time, to make it easiest
to compare to lattice results.

\begin{table}[tbh]
  \centerline{
  \begin{tabular}{|cc|r|r|r|r|} \hline
    & & $\kappa \to \infty$ & $\kappa = 32$
    & $\kappa = 16$ & $\kappa = 6$ \\ \hline
    $n=4$ & $v^2$ &
    0.238 & 0.233 & 0.229 & 0.217 \\
    $(w=0)$ & $\zeta$ &
    11.4 & 10.4 & 9.7 & 8.1  \\ \hline
    $n=8$ & $v^2$ &
    0.159 & 0.158 & 0.156 & 0.151 \\
    $\left( w=-\frac{1}{6}\right)$ & $\zeta$ &
    13.4 & 13.0 & 12.7 & 11.9  \\ \hline
    $n=16$ & $v^2$ &
    0.092 & 0.092 & 0.092 & 0.091 \\
    $\left( w=-\frac{1}{4}\right)$ & $\zeta$ &
    18.7 & 18.6 & 18.5 & 18.2  \\ \hline
  \end{tabular}}
  \caption{\label{table:onescale}
    Expected network velocity-squared $v^2$ and string density $\zeta$
    for local networks and global networks at three values of the tension
    parameter $\kappa$.  The values $n=4,8,16$ correspond to equations
    of state with $w=0,$ $-1/6$, and $-1/4$ respectively.
  }
\end{table}

The table shows smaller network densities and slower strings for the
global cases, especially for the smallest value of $\kappa$.  However,
as the Hubble expansion rate is increased such that the mean network
velocity becomes small, radiation of Goldstone modes becomes
inefficient and the results become almost indistinguishable.
This appears to be a robust prediction of the current one-scale
models.  We will see how these predictions hold up in subsequent
sections.

\section{Networks with fixed, small scale hierarchy}
\label{sec:fixkappa}

In discussing scaling solutions so far, we have ignored the
complication that $\kappa$ is not a constant.  The scale hierarchy at
play is $\kappa \equiv \ln(m/H)$ with $m$ a microscopic scale which is
fixed in physical units, and $H$ the Hubble scale which grows with
time.  This leads to an evolution of the ``scaling'' solution (which
is then not strictly a scaling but a tracking solution).  However, in
a numerical simulation one always starts away from the tracking
solution.  Therefore
previous global string simulations all contain simultaneous evolution
\textsl{towards} the tracking solution, and evolution \textsl{of} the
tracking solution.  Separating these two effects is nontrivial.
Recently the authors of Ref.~\cite{Hindmarsh:2019csc} have claimed that
previous studies are mistaken, and the whole evolution of network
densities in numerical simulations can be ascribed to the approach to
tracking solutions.  Their results are consistent with \textsl{no}
$\kappa$ dependence in the actual tracking or scaling network
solution.  In this section we will show how to perform numerical
simulations of global strings, such that $\kappa$ actually remains a
constant and a scaling solution unambiguously exists.  We can consider
networks at different $\kappa$ values, and we will show that they
unambiguously have different scaling solutions.

Consider cosmic strings arising from a theory of a complex scalar
field
$\varphi = \frac{\phi_r + i \phi_i}{\sqrt{2}}$
with a symmetry-breaking potential.  The Lagrangian is%
\footnote{%
  We use the $[{-}{+}{+}{+}]$ metric convention and natural units
  throughout.}
\begin{equation}
  \label{Lagrangian}
  -\frac{\LL}{\sqrt{-g}} =
  g^{\mu\nu} \partial_\mu \varphi^* \partial_\nu \varphi
  + \frac{m_r^2}{8 f^2} \left( f^2 - 2 \varphi^* \varphi \right)^2
  \,.
\end{equation}
Here $m_r$ is the mass of the radial field excitation (Higgs particle)
and $f$ is the vacuum expectation value of the scalar field.
In a radiation dominated FRW universe and using conformal time $\tau$,
the metric is $g_{\mu\nu} = \tau^2 \eta_{\mu\nu}$,
$g^{\mu\nu} = \tau^{-2} \eta^{\mu\nu}$, and $\sqrt{-g}=\tau^4$.
Therefore, we can rewrite the Lagrangian as
\begin{equation}
  \label{Lagrangian2}
  -\LL = \tau^2 \left( \eta^{\mu\nu} \partial_\mu \varphi^* \partial_\nu \varphi
  + \frac{m_r^2 \tau^2}{8f^2} (f^2-2\varphi^* \varphi)^2 \right) \,.
\end{equation}

Note that the potential term scales as $\tau^2$ relative to the gradient
term.  Therefore, the string core's radius
$r_{\mathrm{core}} \sim 1/m_r$, when expressed in terms of comoving
coordinates, scales as $1/(m_r \tau)$, while the mean
string separation scales as $R \sim \tau$.  Therefore the ratio of
these scales behaves as $R/r_{\mathrm{core}} \propto \tau^2$.  This is
a problem for simulations, because $r_{\mathrm{core}}$ must be
kept larger than the lattice spacing at all times, and the simulation
must end before $R = L$ the box size.

Press Ryden and Spergel \cite{Press:1989yh} (henceforward PRS)
proposed an alternative
numerical approach, in which the $\tau^2$ scaling is simply dropped from
the potential term, such that the string core stays the same size in
comoving coordinates and therefore in lattice units.  While
unphysical, this approach gives the maximum dynamic range over which
the simulation describes an evolving network, and therefore provides a
better chance for the network to approach its tracking behavior.
Some but not all studies take this approach.

But in either of these approaches, the Lagrangian does not display
conformal scaling, and so there is no rigorous argument that the
dynamics will either.  The ratio of string separation to core size
changes with time, which we might expect to lead to an evolution in the
scaled network density.  But at the same time, the initial conditions never
have a string density corresponding to the scaling (or attractor)
solution, so we will \textsl{also} see corrections to scaling arising
from initial conditions.  How do we tell these two corrections apart?

Here we propose to replace $m_r^2 \tau^2$ behavior with $m_r^2/\tau^2$
behavior, that is, to consider a system in which the string core size
grows in comoving coordinates as time increases:
\begin{equation}
  \label{Conformal}
  -  \LL_{\mathrm{proposed}} = \tau^2
  \left( \eta^{\mu\nu} \partial_\mu \varphi^* \partial_\nu \varphi
  + \frac{\lambda}{8 f^2 \tau^2} ( f^2 - 2 \varphi^* \varphi )^2 \right) \,,
\end{equation}
which remains unchanged if we make the substitution
$(x,\tau) \to (\xi x,\xi \tau)$ and
$\varphi(x,\tau) \to \varphi(\xi x,\xi \tau)$.  Therefore the dynamics are
self-similar, in terms of the Hubble scale, as a function of time.
There is every reason to expect this model to approach a scaling
solution at late times, depending only on $\lambda = m_r^2 \tau^2$.
The logarithm of the scale hierarchy is given by
$\kappa = \ln(\sqrt{\lambda})$.  We can then investigate network
dependence on the core-to-separation hierarchy by performing
simulations with a range of $\lambda$ values.

\begin{figure}[htb]
  \centerline{
  \includegraphics[width=0.65\textwidth]{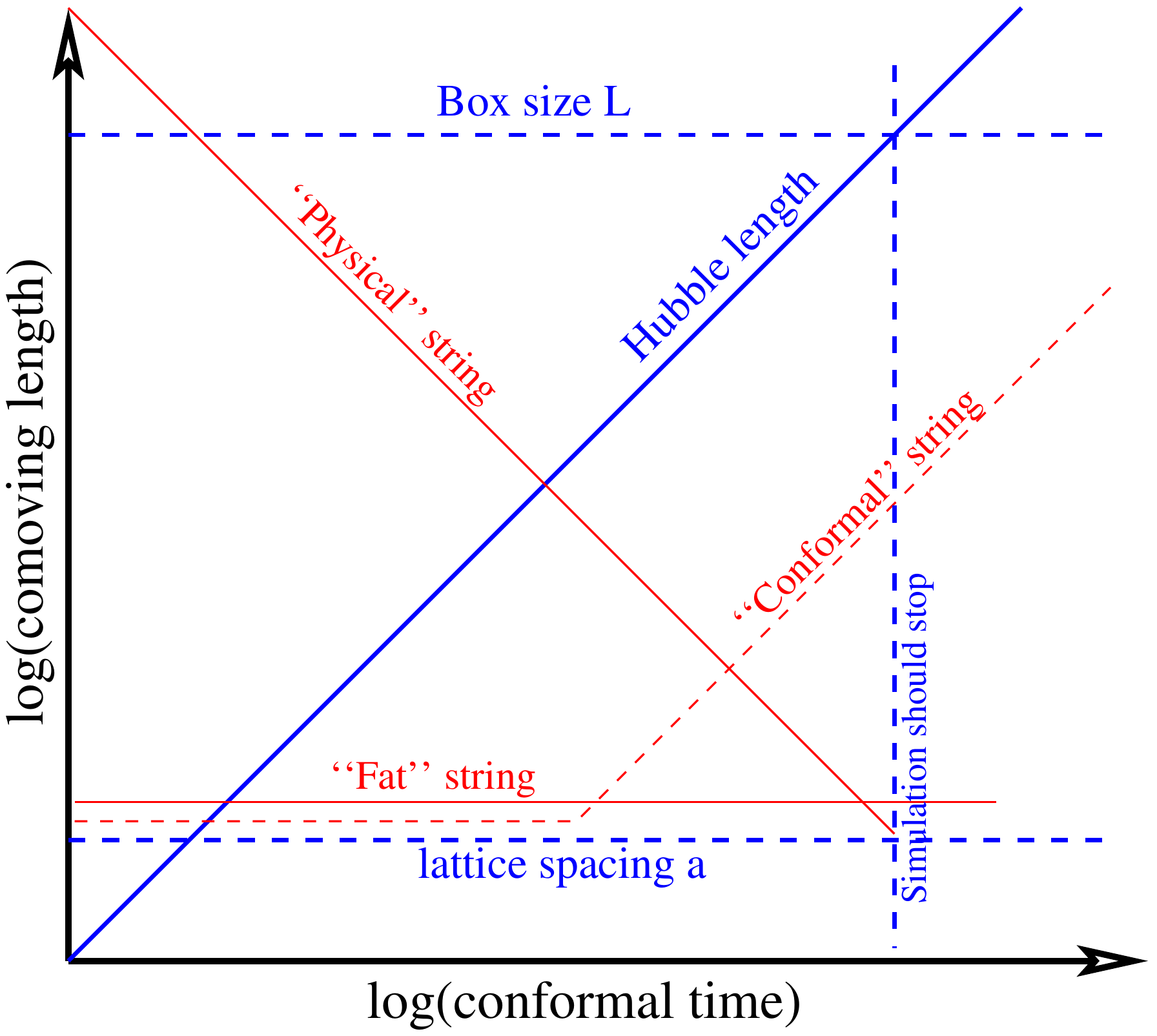}}
  \caption{\label{fig_logscales}
    The plot shows different ways the log of a comoving scale
    (vertical axis) can depend on the log of conformal time
    (horizontal axis).  The Hubble scale is a straight line of slope
    1.  The lattice spacing and lattice length scales are horizontal
    lines.  The physical case, where the string core is of fixed size
    in physical units, is a downward-sloping line, while the
    PRS or ``fat string'' case is a straight line.  We
    propose to follow a straight line until a given $m_r/H$ ratio is
    achieved, and then to follow a slope-1 line, keeping
    $\ln(m_r/H)$ fixed -- conformal scaling, allowing a true attractor
    solution.}
\end{figure}

Figure \ref{fig_logscales} provides a hopefully helpful cartoon of
these three possibilities.  In a log-log plot of the comoving length
against the conformal time, the lattice spacing $a$ and the lattice
extent or length $L$ are flat lines, while the Hubble scale $H^{-1}$ is a
diagonal line of slope 1.  The physical case has a string core size
which falls (in comoving units and in terms of comoving time)
as $t^{-1}$.  The network only really comes into existence once
$m_r/H \gg 1$, so the dynamic range available (the interval of time
during which the curve is below the Hubble curve but above the
lattice-spacing curve) is compressed.  The PRS
approach keeps the core size fixed in lattice units, which provides
the maximum possible dynamic range.  Our proposal initially follows
the PRS choice, but then switches to a slope-1 line,
that is, $m_r \propto t^{-1}$, so the ratio $m_r/H$ remains fixed, the
theory becomes conformal, and a true scaling solution exists.

In carrying out a numerical study we make the following
modifications.%
\footnote{In other respects our numerics are rather standard, eg,
  cubic $N\times N\times N$ box with periodic boundary conditions,
  Leapfrog update algorithm, $\varphi$ initialized to lie on the
  vacuum manifold with independent phase at each point in space.  We
  approximate the $\nabla^2$ term in the equations of motion with an
  improved next-nearest neighbor approximation
  (13-point rather than 7-point stencil).}
First, at early times \Eq{Conformal} calls for a mass which is large
compared to the inverse lattice spacing.  Instead, we start with
$m_r = 1.5/a_x$ ($a_x$ the lattice spacing) and we switch over to
$m_r = \sqrt{\lambda}/\tau$ once this quantity falls below $1.5/a_x$.
Second, we increase the Hubble damping strength by a factor of 8
(strongly damped early evolution)
until $\tau=24a_x$ so that the initial string density will be higher,
close to the scaling density.  In treating the resulting data we will
only make use of that part of the simulation which occurs well after
these modifications have been switched off.

We performed simulations with
$m_r \tau = 20,$ $50$, $100$, $200$, and $500$; the first three on a
$1024^3$ lattice to $\tau = 512 a_x$ and the latter two on a
$2048^3$ lattice to $\tau = 1024 a_x$ (where $a_x$ is the lattice
spacing).  The final time is chosen so that
causality still ensures the results to be equivalent to the
infinite-volume limit.  For the smaller volumes we gather 32
independent evolutions; for the larger volumes we gather 8.
In each simulation we evaluate the
length of string by counting plaquettes pierced by a string and then
multiplying by 2/3 to correct for the direction-average of the
number of plaquettes pierced per unit length of string (Manhattan
effect \cite{Press:1989yh,Scherrer:1997sq}).  We also
estimate the gamma-factor of the string, using the method of
Fleury and Moore%
\footnote{%
  The method uses the value of $|\partial_\tau \varphi|^2$ at the points
  around each plaquette which the string pierces.  For the known
  profile of a straight string, this can be used to determine
  $v^2/(1-v^2)$ at this point on the string.  The method receives
  errors when $(ma)$ is not large enough, which become small late in a
  simulation.  It is also systematically wrong on highly curved parts
  of string, parts with large ``breather'' mode fluctuations, or near
  string intersections.  These issues are expected to diminish with
  increasing $m_r \tau$.}
\cite{axion1}.  We then report the total string
length, scaled by factors of the time and the lattice volume such that
it should scale, both with and without a factor of $\gamma$:
\begin{equation}
  \label{zetadef}
  \zeta_{\mathrm{len}} \equiv \frac{\ltot \tau^2}{V}
  = \frac{\frac{2}{3} n_{\mathrm{pierced-plaq}} (\tau/a_x)^2}{N^3} \,, \qquad
  \zeta_{\mathrm{inv}} \equiv \frac{\ltotg \tau^2}{V}
  = \frac{\frac{2}{3} (\tau/a_x)^2 \sum_{\mathrm{pierced-plaq}}\gamma}{N^3}
  \,.
\end{equation}
Note that we are reporting the network density in comoving coordinates
and conformal time, as we do throughout this paper.
The factor $\gamma$ is conventionally included when finding the
network density because it accounts for the energy content, rather
than comoving-frame length, of the string.

\begin{figure}[tbh]
  \includegraphics[width=0.47\textwidth]{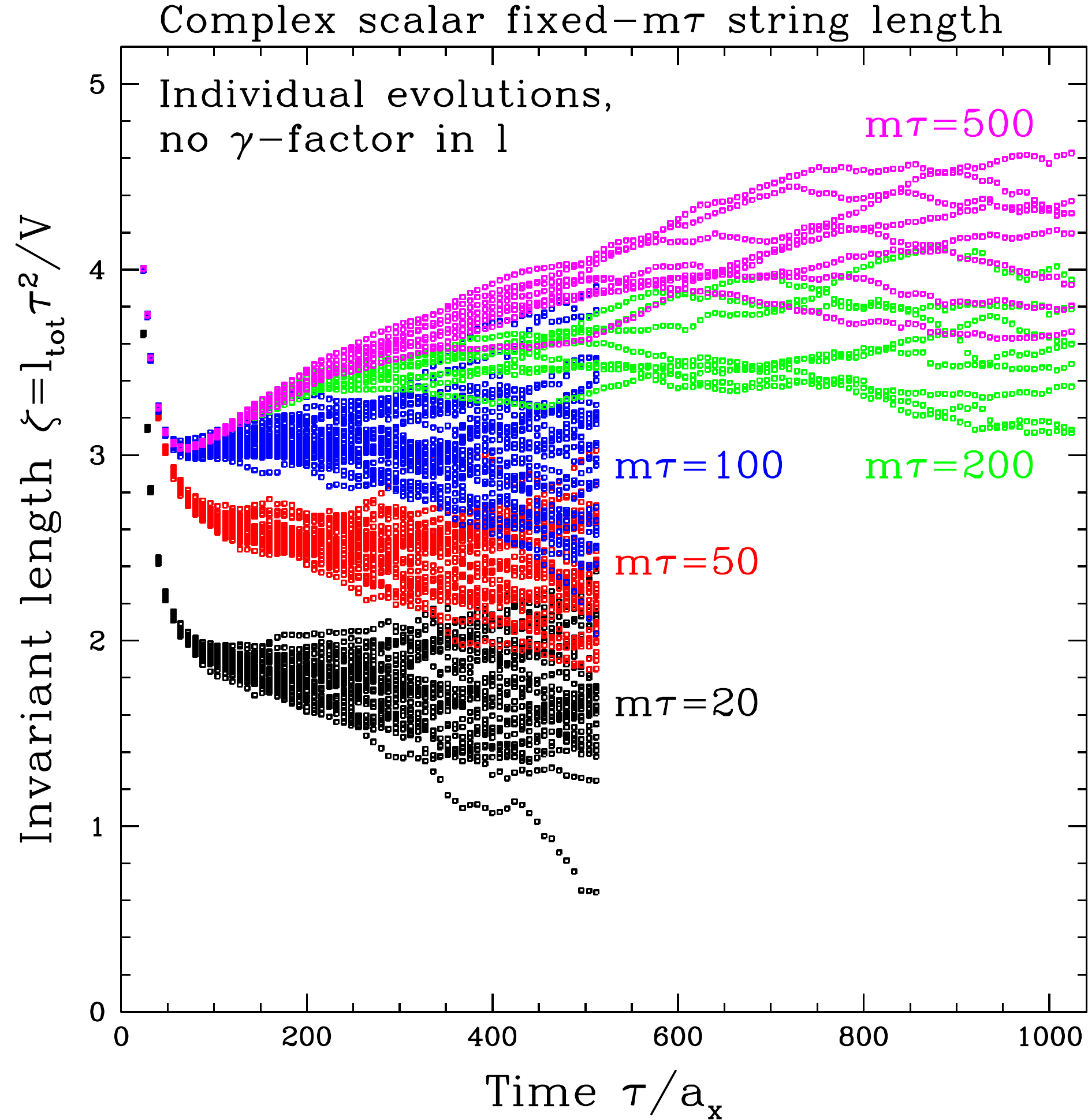}
  \hfill
  \includegraphics[width=0.47\textwidth]{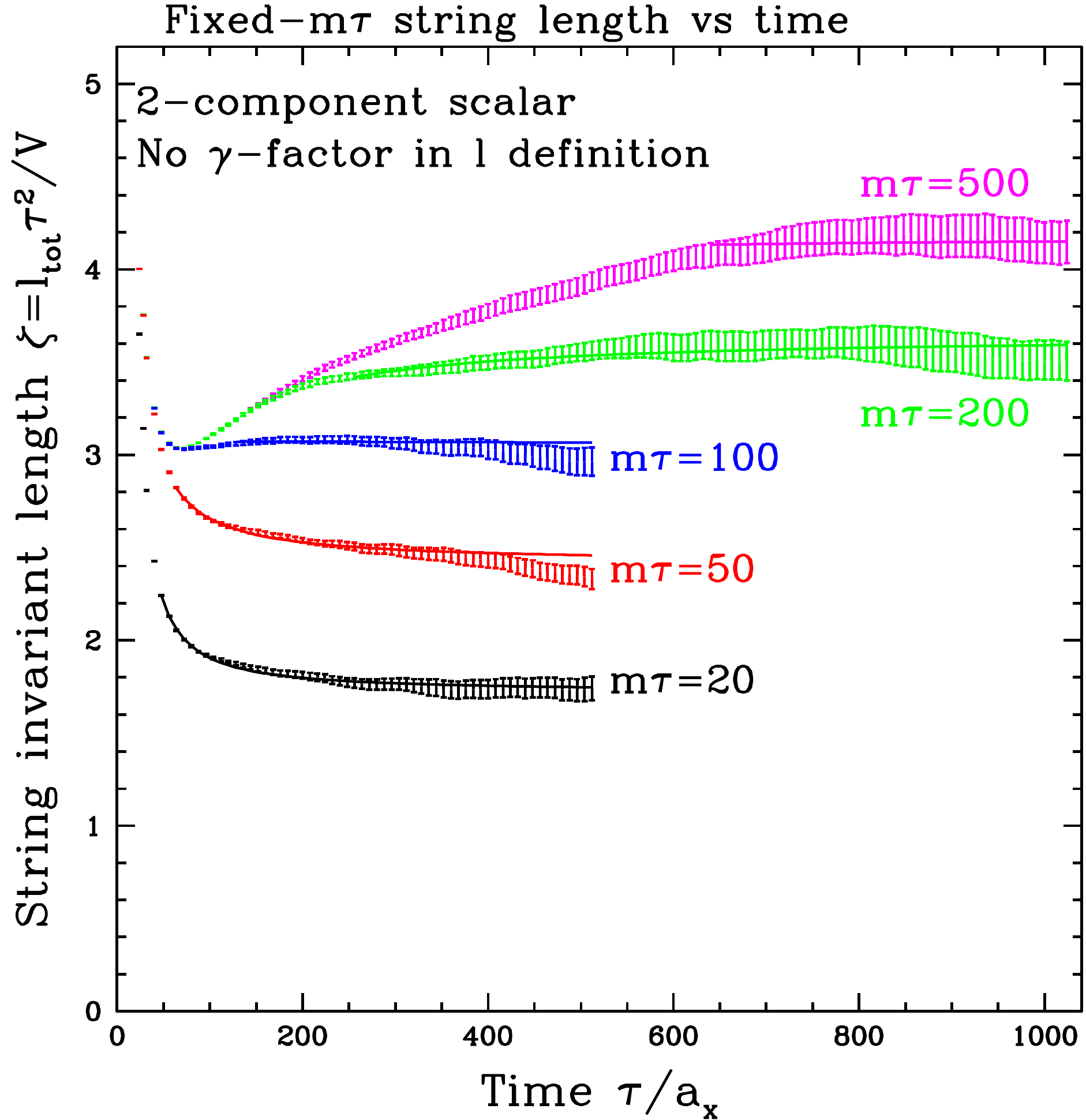}

  \vspace{2ex}
  
  \includegraphics[width=0.47\textwidth]{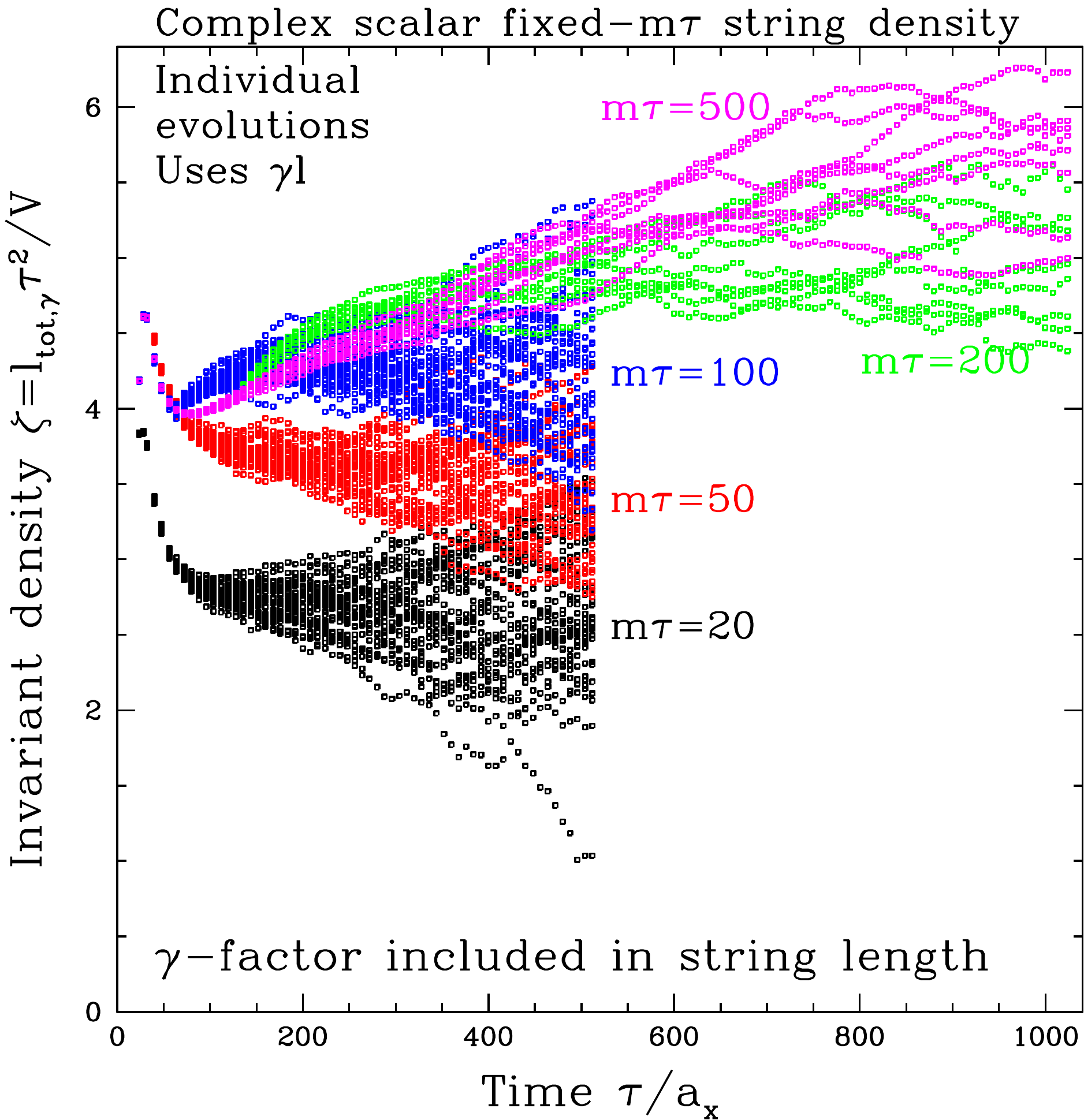}
  \hfill
  \includegraphics[width=0.47\textwidth]{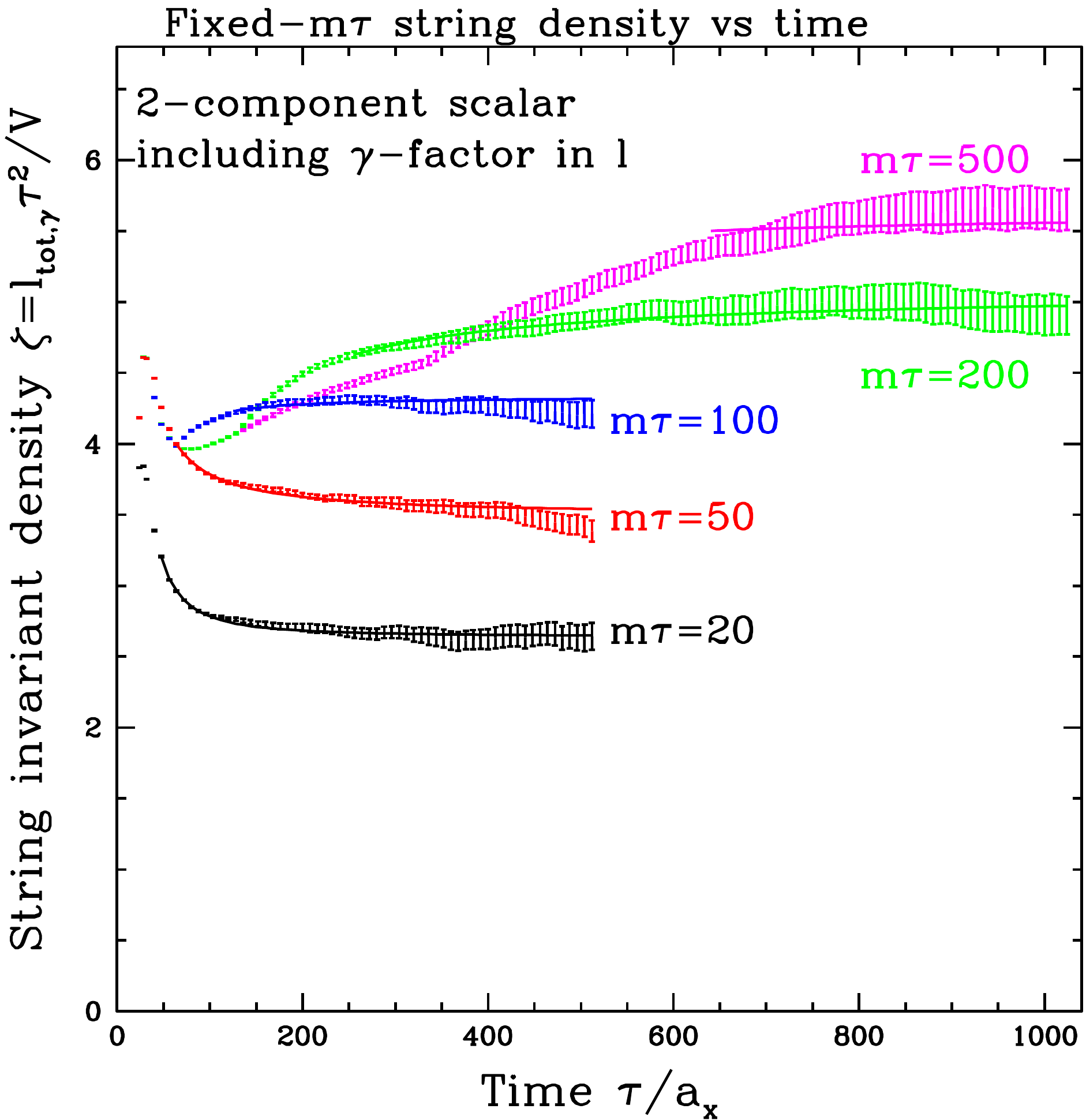}  
  \caption{\label{fig:fatstring}
    Raw data (left) and average over simulations (right) with
    $1\sigma$ statistical error bars, for the
    scaled string length at several $m_r \tau$ values.  The top figures
    use the geometrical string length, the bottom figures include the
    string gamma-factor in computing the string length.  Right-hand
    figures also show the best-fit lines.}
\end{figure}

Figure \ref{fig:fatstring} shows the raw and averaged data for each
$m\tau$ value we consider.  We see very clearly that each evolution
rapidly approaches a plateau, and that the plateau value shows clear
dependence on $m\tau$, with larger $m\tau$ (smaller core size) giving rise
to a denser string network.  The figure also displays our fit to the
data. We use the time range during which $m a_x < 0.78$, and we fit
including coefficients for
$\tau^{-1}$ and $\tau^{-2}$ early-time transients, which we weakly constrain
with priors that they are initially of order $20\%$ and $4\%$ of
the total network density.  The resulting best-fit network densities
are summarized in Table \ref{tab:fatstring}.  The table also reports
the mean-squared string velocity
$\int \gamma v^2 d\ell / \int \gamma d\ell$.  Here we use the
Fleury-Moore estimate for velocity and evaluate at the time when
$ma_x=0.5$, such that the lattice-spacing errors are comparable for each
case.  We indicate the statistical errors only; by considering twice
smaller values of $ma_x$ (twice-later times) we find the errors in
$\langle v^2 \rangle$ from lattice spacing and transient effects are
$\sim 0.005$.  The
systematic errors from, eg, breather modes, string curvature, and
exactly what we mean by the velocity of a string with finite
core thickness are difficult to estimate but are the most severe for
the smallest $m_r \tau$ values.

\begin{table}[tbh]
\centerline{  \begin{tabular}{|r|r|r|r|r|}
    \hline
    $m_r t$ & $\kappa\;$ & $\zeta_{\mathrm{len}} \hspace{1.2em}$
    & $\zeta_{\mathrm{inv}}\hspace{1.2em}$ &
    $\langle v^2 \rangle \hspace{1.2em}$ \\
    \hline
    20 & 3.0 & $1.74(3)\po$ & $2.64(3)\po$ & $0.4919(3)$ \\
    50 & 3.9 & $2.42(2)\po$ & $3.49(3)\po$ & $0.4731(5)$ \\
    100 & 4.6 & $3.06(4)\po$ & $4.34(5)\po$ & $0.446(1)\po$ \\
    200 & 5.3 & $3.65(10)$ & $5.08(12)$ & $0.423(3)\po$ \\
    500 & 6.2 & $4.18(11)$ & $5.65(14)$ & $0.419(7)\po$ \\
    \hline
  \end{tabular}}
  \caption{\label{tab:fatstring}
    String density, with and without $\gamma$ factor, as a function of
    the (constant) $mt$ value during the simulation, after
    extrapolation to late time; and mean squared string velocity.}
\end{table}

\begin{figure}
  \centerline{
    \includegraphics[width=0.65\textwidth]{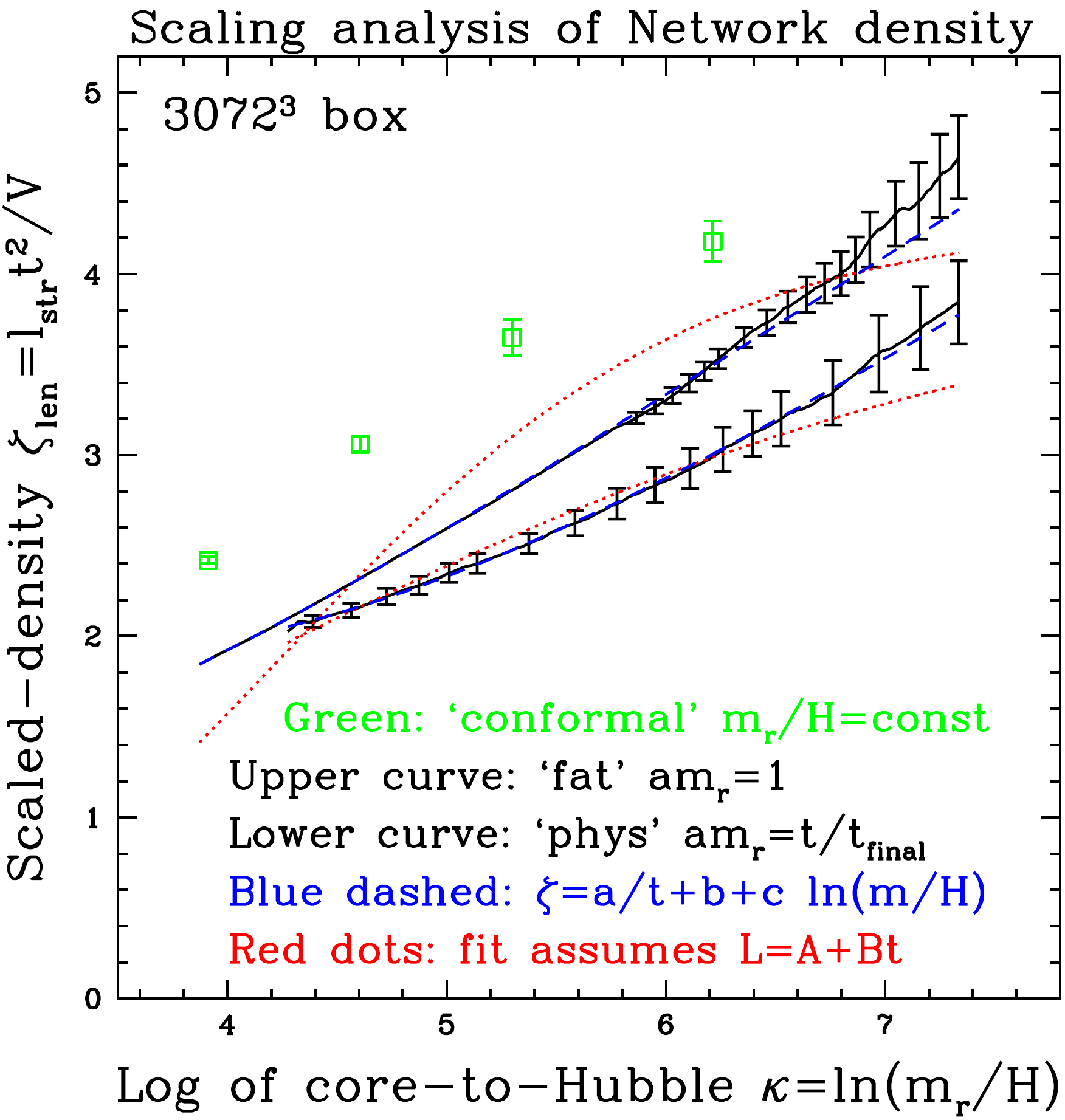}}
  \caption{\label{fig:xievolve}
    network density $\zeta_{\mathrm{len}}$ against $\kappa$ for conformal
    networks (green points), PRS networks (upper curve, labeled as
    ``fat'') and physical-string networks (lower curve)
    The PRS and physical-string network
    densities lag the conformal-network value as expected.  Blue and red
    fits are explained in the main text.}
\end{figure}

How do we expect this data to be related to the
cases of the PRS network or the ``physical'' network?
Call the density of the conformal network with a given
$\kappa=\ln(m_r/H)$ value $\zeta_c(\kappa)$.
Intuitively, at a given moment when $\kappa(t)=\ln(m_r(t)/H(t))$ takes
a given value, the string network density $\zeta(t)$ should evolve
towards the attractor value $\zeta_c(\kappa(t))$; for small
differences, we expect
\begin{equation}
  \label{zeta-evolves}
  \frac{t\, d\zeta(t)}{dt} = C (\zeta_c(\kappa(t)) - \zeta(t))
\end{equation}
with $C$ some order-1
constant, telling how fast a network approaches the scaling behavior.
But for PRS or ``physical'' networks, we have
$d\kappa/d\ln(t) =1$ (PRS) or 2 (``physical'').  Therefore,
$\zeta(t)$ is evolving towards a moving target.
If $\zeta_c(\kappa)$ depends approximately linearly on $\kappa$,
$d\zeta_c/d\kappa = \zeta'$,
then the ``tracking'' solution to \Eq{zeta-evolves} is
$\zeta(t) = \zeta_c(\kappa(t)) - \frac{\zeta'}{C} \times (1\mbox{ or }2)$.

We can check this picture by simulating PRS networks and physical
networks and plotting the resulting $\zeta(\kappa(t))$ on the same plot as
$\zeta_c(\kappa)$.  Our results are shown in Figure \ref{fig:xievolve},
which shows the network density $\zeta_{\mathrm{len}}$ as a function of
$\kappa=\ln(m_r/H)$ for PRS and physical networks, carried out on
$3072^3$ boxes up to time $t=1536a$ with $m_r=1/a$ in the PRS case and
$m_r = t/t_{\mathrm{final}}$ in the physical case.  In each case the
field is initialized to a random independent value at each point and
then evolved under radiation-era damping without an initial
high-damping stage.  The blue curves are fits to an assumed behavior
with an early-time transient of form $a/t$ and a linear dependence on
$\kappa$.  The red curves are fits following the assumption of
\cite{Hindmarsh:2019csc} that the correlation length of the network,
$L \equiv \sqrt{V/\ltot}$, scales as $L=A+Bt$, that is, linearly in
time but with an offset.  This does not provide a good description of
the data over the full range shown.
The figure indeed illustrates that the PRS network is shifted down (or
rightward) with respect to the conformal scaling solutions, and the
``physical'' network is shifted by a larger coefficient, which within
errors and initial transients is consistent with a factor of 2.

In summary, our data shows robustly that the network density is
sensitive to the core-to-separation ratio of the network.  Both
lattice spacing effects and initial conditions are thoroughly under
control in this determination.  The standard PRS and ``physical''
networks appear to evolve towards tracking solutions which trail
behind the scaling solutions we find in the conformal network case.

It would be straightforward to extend this treatment to larger damping
strengths $n>2$.  But instead we will consider a wider range of string
tensions at high damping, in the next section.

\section{Networks with large tension}
\label{subsec:scaling}

Realistic networks have $\kappa \geq 70$ as noted before, so
simulations of U(1) symmetric scalar field theory are quite far
from the regime where they reproduce the physically relevant network
dynamics.  Therefore we need some method to study
networks with much larger $\kappa$ values.

In \cite{axion3} we present a way to do so,
based on effective field theory ideas from \cite{Dabholkar:1989ju}.
Consider a network with an extremely large scale hierarchy,
$1/m \ll t$ or $1/H$.  Then we can ask what the network looks like,
after ``fuzzing'' details smaller than an intermediate scale
$r_{\mathrm{cut}} \ll 1/H$ but $r_{\mathrm{cut}} \gg 1/m$.  On this
scale strings are nearly straight and have negligible core thickness.
The string then looks like an object with tension
$T = \pi f^2 \ln(r_{\mathrm{cut}} m)$, interacting with long-range
Goldstone modes via a Kalb-Ramond interaction \cite{Dabholkar:1989ju}.
If we find some other -- any other -- model which also behaves, at the
scale $r_{\mathrm{cut}}$, like strings with tension $T$ interacting
with Goldstone modes via a Kalb-Ramond interaction, then it will
describe the evolution of the same combination of string network plus
Goldstone fields.
We refer the reader to our previous paper \cite{axion3} for further
details.  Suffice it to say that we can add a constant extra string
tension to the string core, $\kappa \to \kappa + \Delta \kappa$, and
that the most convenient values for $\Delta \kappa$ are of form
$2(i^2 + (i+1)^2)$ with $i$ an integer, for instance $10$ ($i=1$),
$26$ ($i=2$), $50$ ($i=3$), and so forth.  This is in addition to the
logarithm $\ln(m\tau)$ with $m$ the mass associated with the string core
(which should obey $ma_x \leq 1$ with $a_x$ the lattice spacing, so the
core is resolved by the lattice), and with $\tau$ the conformal time
which sets the inter-string separation (IR) scale.  Since our lattices
are $2048^3$ and we use the time range $\tau \in [512a_x,1024a_x]$, we have
approximately $\kappa = 6 + \Delta \kappa$.  We will perform
simulations of the pure complex scalar theory, the modified theory
with $\Delta \kappa = 10$ and $\Delta \kappa = 26$, and the abelian
Higgs model.  In what follows we will refer to these cases as
$\kappa = 6$ (simulations with complex scalars only),
$\kappa = 16$ (simulations with two scalars and a gauge field, with
charges $2,1$),
$\kappa = 32$ (simulations with two scalars and a gauge field, with
charges $3,2$), and
$\kappa \to \infty$ (abelian-Higgs model).
We perform a few simulations for each combination, using the
fluctuations between simulations -- and between a larger number of
simulations on $1024^3$ lattices -- to estimate statistical errors.  Because we
use an improved action, we choose the mass scale for all heavy
excitations to be $ma_x=1$ for each case.  Previous results show that
making the lattice spacing finer makes little difference in the
results \cite{axion1,axion3}.

In this section we will only consider $n \geq 4$, that is, strong Hubble
damping corresponding to an equation of state with $w \leq 0$.  In
particular we study $n=4$, $n=8$, and $n=16$.
This parameter enters in the dynamics, in conformal coordinates,
through an overall coefficient on the Lagrangian; \Eq{Conformal} is
replaced with
\begin{equation}
  \label{Lagrangian:n}
  \LL = \tau^n \left( \eta^{\mu\nu} D_\mu \varphi^* D_\nu \varphi
  + \frac{m^2}{8f^2} (f^2 - 2 \varphi^* \varphi)^2
  + \frac{1}{4e^2} \eta^{\mu\alpha} \eta^{\nu\beta} F_{\mu\nu} F_{\alpha\beta}
  \right) \,,
\end{equation}
where for the pure scalar case $D_\mu \to \partial_\mu$ and for the
two-scalar case (to get intermediate string tensions) the scalar term
should be duplicated with two fields with different charges as
described in \cite{axion3}.  Note that we use the Press-Ryden-Spergel
method for both the scalar potential and the gauge fields, such that
both gauge and scalar masses remain fixed in comoving coordinates.
When deriving equations of motion, the factor $\tau^n$ leads to
\begin{equation}
  \label{n-role}
  \partial_\tau^2 \varphi \rightarrow
  \partial_\tau^2 \varphi + \frac{n}{\tau} \partial_\tau \varphi \,,
  \qquad
  \partial_\tau E_i \rightarrow \partial_\tau E_i + \frac{n}{\tau} E_i \,,
\end{equation}
such that $n$ controls the strength of dissipative forces in the
field evolution and therefore in the network's evolution.

We already explained the rationale for studying disparate $n$ values,
in terms of probing a broader range of network
behaviors.  But there is a second reason, which is that numerical
simulations become better under control as $n$ gets larger.  This is
because Hubble damping tends to smooth the strings and remove the
locations with the largest velocities; short-distance structures and
fast-moving (highly Lorentz contracted) strings are two things which
simulations handle badly, so the simulations become more faithful as
we move in this direction.  Simulations tend to approach the scaling
network density from below, and we accelerate this approach by
starting our simulations with a much larger Hubble damping rate, so
that the network initially evolves slowly and remains dense; this
high-damping regime is turned off around $\tau=128a_x$, and
the strength of early damping is tuned such that the network density
at $\tau=256 a_x$ and $\tau=512 a_x$ are the same.  We
will see that the network is nevertheless not quite in the scaling
limit; the network density increases somewhat between $\tau=512a_x$ and the
end of the simulation at $\tau=1024a_x$, leaving some non-scaling
corrections in our results.  But the difference between different
network types will be much larger than these non-scaling corrections,
so we feel that the simulations still provide interesting results.

\begin{figure}
	\begin{subfigure}{0.49\textwidth}
		\includegraphics[width=\textwidth]{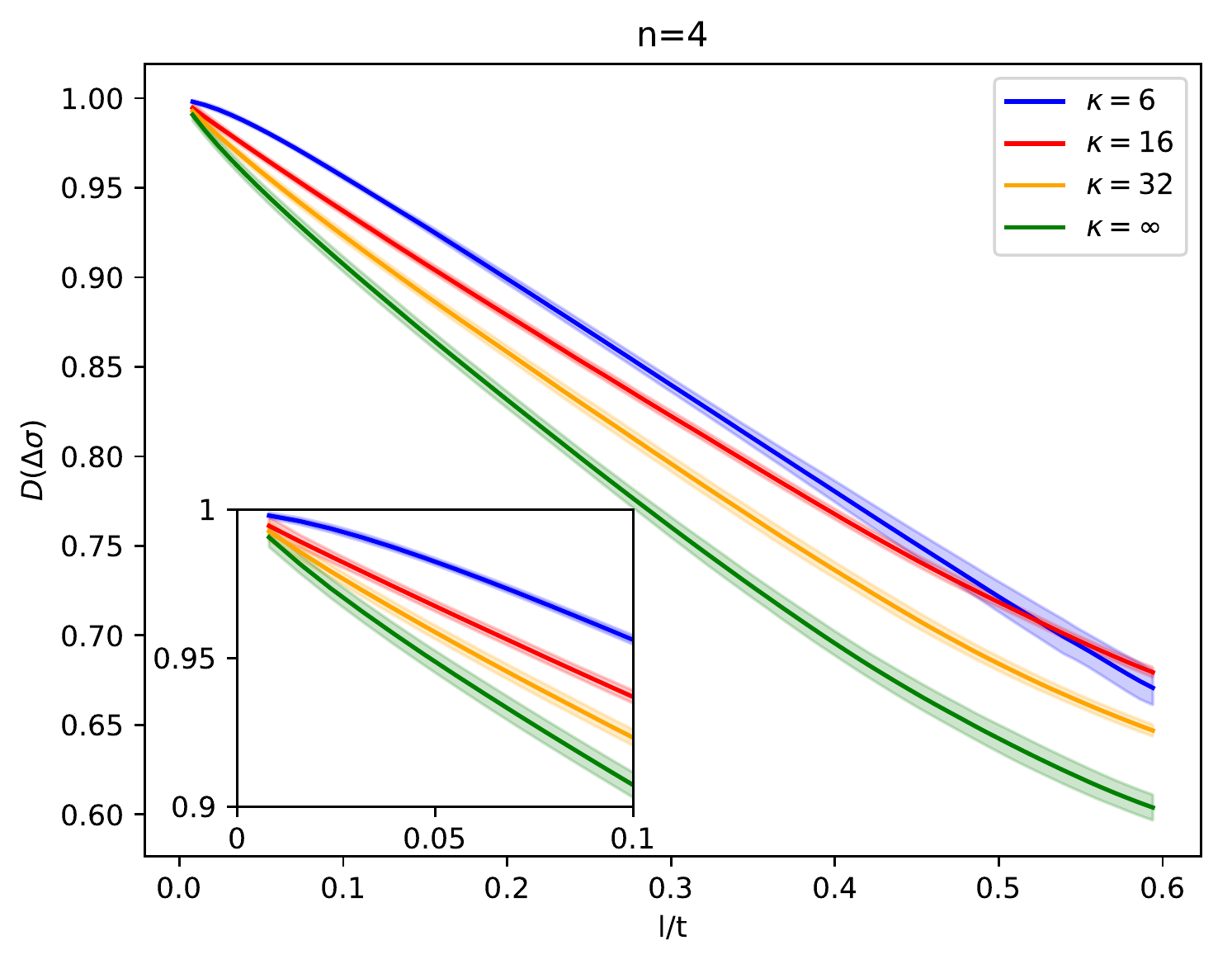}
		\caption{}
	\end{subfigure}
	\begin{subfigure}{0.49\textwidth}
		\includegraphics[width=\textwidth]{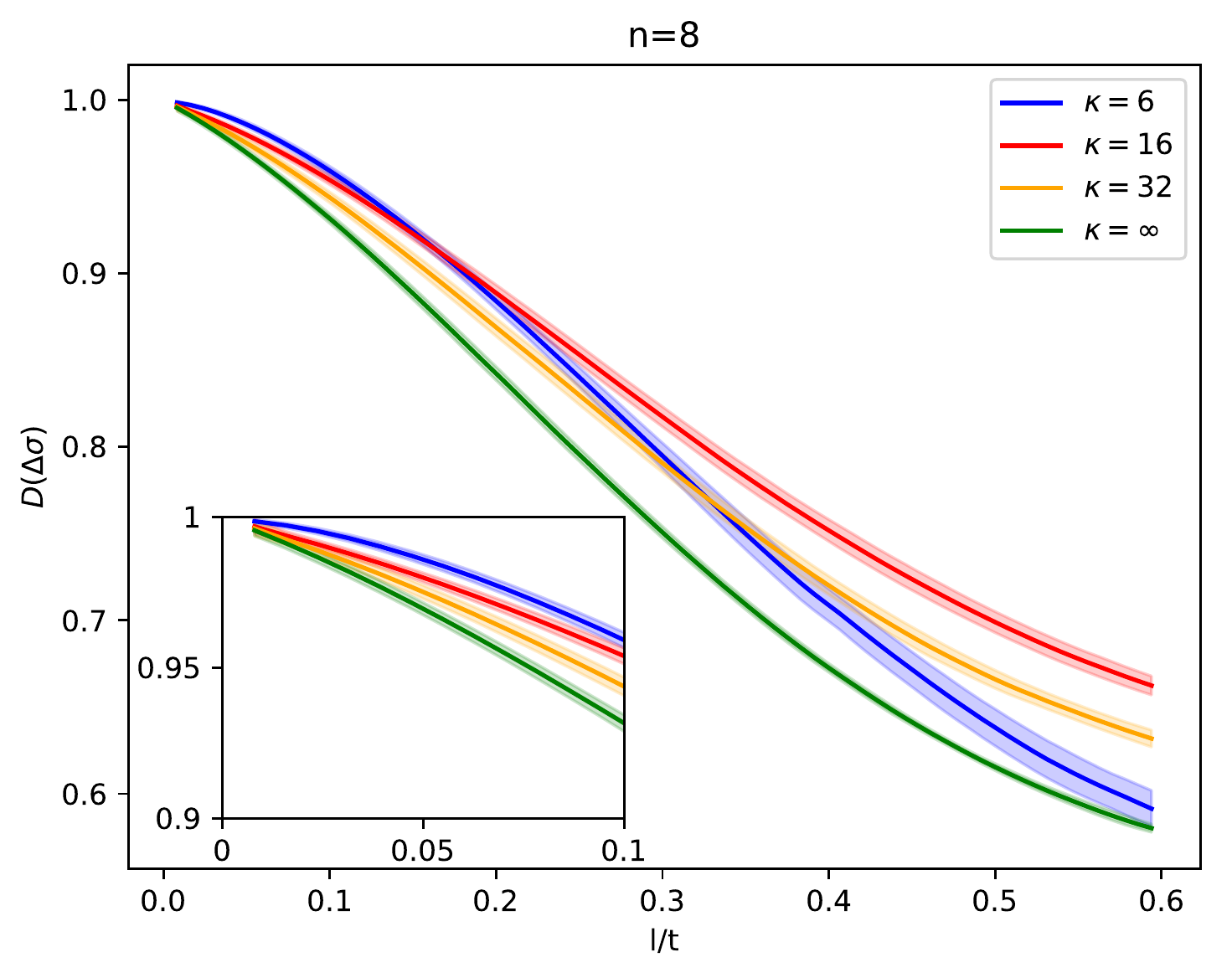}
		\caption{}
	\end{subfigure}
        \centerline{
	\begin{subfigure}{0.49\textwidth}
		\includegraphics[width=\textwidth]{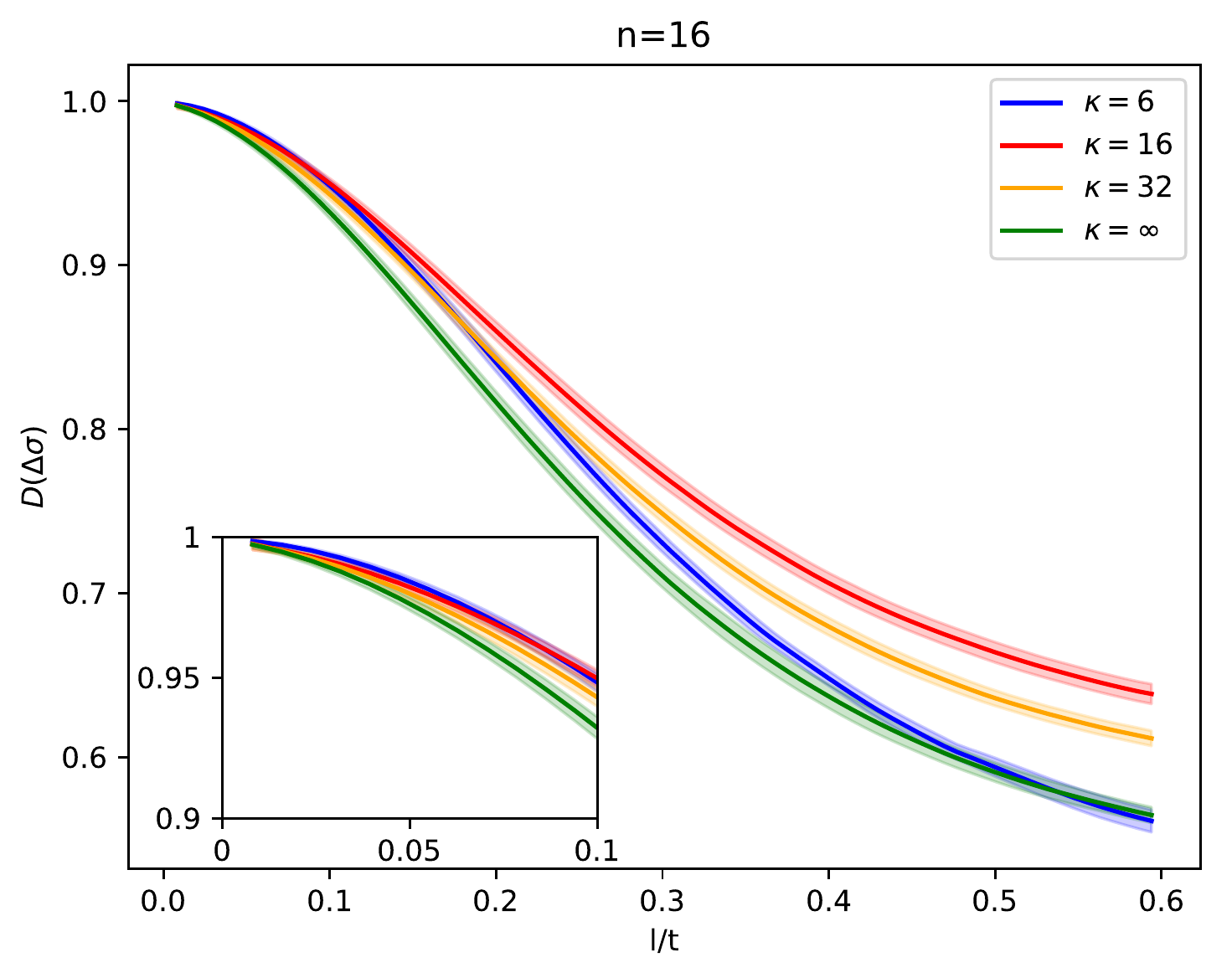}
		\caption{}
	\end{subfigure}}
  \caption{\label{fig:angleauto}%
    The string-tangent autocorrelation function for strings with four
    different tensions $\kappa$, at three Hubble drag strengths
    $n=4$ ($w=0$), $n=8$ ($w=-1/6$), and $n=16$ ($w=-1/4$).}
\end{figure}

We start by examining whether the networks really have reduced
small-scale structure as $n$ is increased.
We do so by examining the string-tangent autocorrelation
function as a function of separation along a string.  That is, at each
point on the string we can define the string unit-tangent direction
$\hat x'(\sigma)$ with $\sigma$ the affine parameter indicating the position
along the string.  The autocorrelation function is
\begin{equation}
  \label{tangentauto}
  D(\Delta \sigma) \equiv \frac{1}{\ltot} \int_0^{\ltot}
  \hat{x}'(\sigma) \cdot \hat{x}'(\sigma{+}\Delta \sigma)
  \;  d\sigma \,.
\end{equation}
This tells how quickly the string's direction changes as one moves
along the string.  Details of how we determine the string's location
are provided in Appendix \ref{AppA}.  The autocorrelator is shown in
Figure \ref{fig:angleauto}.  As with all studies of string properties,
the figure is based on an average over \textsl{all} string in a
simulation, both long strings and short loops; however the long
strings dominate the network, so removing loops before taking averages
would lead to modest changes.

The most pertinent feature seen in the figure is that, as the damping
gets stronger, the behavior of the autocorrelator near zero goes from
being approximately linear to being approximately quadratic.  A linear
behavior reflects a string with kinks -- points of abrupt change in
the unit tangent -- while a quadratic behavior indicates a string
which is locally smooth with finite radius of curvature.  We see that
this difference applies equally well for local strings -- the damping
rounds off the kinks.  Kinkless strings are expected to be well
described in field-theoretical simulations.  Therefore, while field
theory simulations are probably not reliable at reasonable lattice
size and spacing for radiation-era ($n=2$) simulations, they may be
reliable for $n \geq 8$.
The figure also shows that small $\kappa$ corresponds to more rounded
strings and longer-distance directional correlations along the
string.  The exception is that, for $\kappa = 6$, while the direction
is highly correlated over short distances, it becomes less correlated
at larger distances, falling below some other curves.  We do not yet
understand this effect.

Next we examine the string's mean squared velocity.  We determine this
velocity in two ways.  The first is from the rate of variation of the
fields at the string's core, as described in \cite{axion1,axion3}.  The
second is geometrical.  We identify the set of points the string goes
through by interpolating the crossing-point within each
plaquette penetrated by the string, as explained in Appendix
\ref{AppA}.  The string is taken as the set of straight line segments
connecting these plaquette-crossing points.  Then we compare all
points on the string network at time $t+\Delta t$ to the network at
time $t$, finding the closest point on the network at time $t$ and
taking the velocity to be distance over time.  We rescale values
$v > 0.95$ back to $v=0.95$ to prevent superluminal motion (which can
occur close to cusps or string intersections).  We have used
$\Delta t = 2a_x$, but we check that the answers are almost the same
using $\Delta t = 4a_x, 6a_x,$ and $8a_x$.

\begin{figure}
	\centering
        \includegraphics[width=\textwidth]{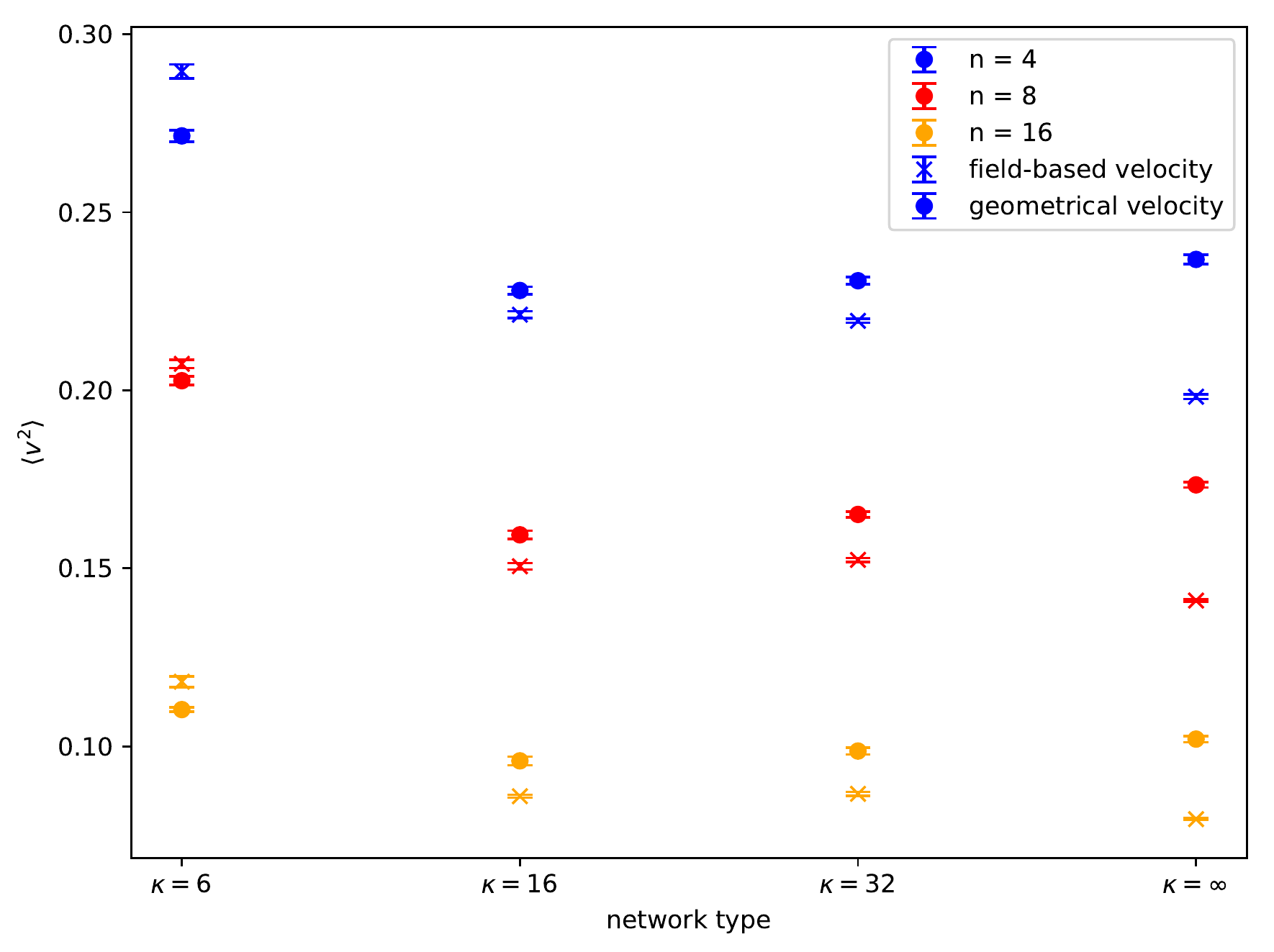}
        \caption{  \label{fig:velocity}
          Mean-squared string velocity, measured two ways:
          using the geometrical definition (measured distance between
          strings at different time steps); and using the
          field-derivative method \cite{axion1}.  The string velocity
          decreases as we increase the Hubble damping strength, and it
          is higher for low-tension scalar-only networks than for
          networks with larger string tension.}
\end{figure}

Our results are presented in Figure \ref{fig:velocity}.  The figure
shows that the mean squared velocity gets smaller as we increase the
strength of Hubble damping, as expected.  There is some discrepancy
between the two estimates of the string velocity, but it is clear from
both methods that the scalar-only simulation provides the largest
string velocity, with higher-tension networks displaying a smaller
mean velocity.

\begin{figure}
	\includegraphics[width=\textwidth]{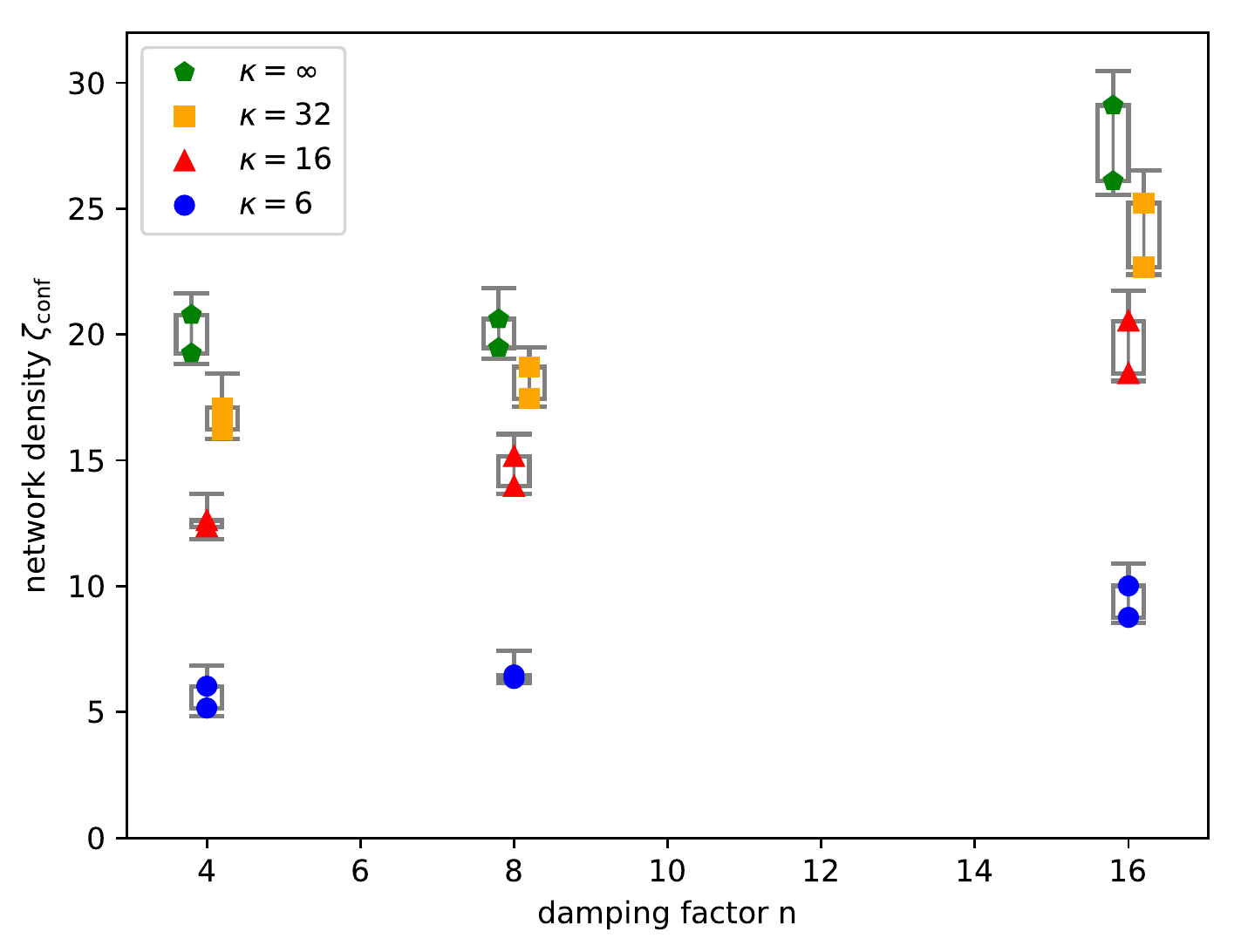}
  \caption{\label{fig:zeta}
    Network density for global and local string networks with various
    tensions at three Hubble-damping strengths.  In each case the
    lower/upper data point is the network density at $\tau=512a_x$ /
    $\tau=1024a_x$, with the lower/upper extending error bar the
    statistical error for that point.}
\end{figure}

Next we turn to the network density.  We define as before the
invariant network density
\begin{equation}
  \label{zeta-recap}
  \zeta_\mathrm{conf} \equiv \frac{\tau^2}{\xi^2} =
  \frac{\tau^2 \int \gamma
    d\ell_{\mathrm{comoving}}}{V_{\mathrm{comoving}}} .
\end{equation}
We have again included a factor of the local
$\gamma$-factor in the relation between the string length and the
network density.  We average this quantity between simulations, once
at time $\tau=512 a_x$ and again at time $\tau=1024 a_x$, to find any residual
corrections from the approach to scaling.  The results are shown in
Figure \ref{fig:zeta}.  We see immediately that the network density
increases as we increase the damping strength.  We also find that
the network density is systematically larger as we consider
higher-tension networks -- or perhaps more accurately, systematically
smaller as we consider networks with stronger interactions with the
Goldstone modes.  This persists at large Hubble damping, in strong
contrast to the expectations from one-scale models.  Considering
abelian-Higgs networks, we find a somewhat larger network density than
\cite{Correia:2019bdl}, probably because our boxes are a factor of 4
larger than theirs, so finite core-to-separation ratio effects are
less under control in their simulations. This might imply that our
results would change further if we had access to larger boxes; indeed
we see a difference between $\tau=512a_x$ and $\tau=1024a_x$ which
indicates that this would be the case.

\begin{table}[ht!]
  \centering
  \begin{tabular}{|c  c | c | c | c | c|}
    \hline
    \multicolumn{2}{|c|}{Network} & $\zeta_\mathrm{conf}$ &
    $\langle v^2 \rangle_{\text{fb}}$ &
    $\langle v^2 \rangle_{\text{g}}$ & $c_{\text{loop}}$\\
    \hline
    & n = 4 & \SI{5.16 \pm 0.33}{} $\rightarrow$ \SI{6.03 \pm 0.81}{}
    & \SI{0.2896\pm0.0019}{} & \SI{0.2715\pm0.0016}{}
    & \SI{0.658\pm0.039}{}\\
    $\kappa = 6$& n = 8
    & \SI{6.33 \pm 0.17}{} $\rightarrow$ \SI{6.47 \pm 0.96}{}
    & \SI{0.2074\pm0.0012}{} & \SI{0.2027\pm0.0012}{} & \SI{0.231\pm0.024}{}\\
    & n = 16 & \SI{8.76 \pm 0.224}{} $\rightarrow$ \SI{10.01 \pm 0.89}{}
    & \SI{0.1181\pm0.0015}{} & \SI{0.1103\pm0.0006}{} &
    \SI{0.0395\pm0.0077}{}\\
    \hline
    & n = 4 & \SI{12.36 \pm 0.49}{} $\rightarrow$ \SI{12.61 \pm 1.05}{}
    & \SI{0.2212\pm0.0010}{} & \SI{0.2280\pm0.0010}{}
    & \SI{0.0622\pm0.0075}{}\\
    $\kappa = 16$& n = 8
    & \SI{13.97 \pm 0.31}{} $\rightarrow$ \SI{15.15 \pm 0.88}{}
    & \SI{0.1506\pm0.0009}{} & \SI{0.1594\pm0.0012}{}
    & \SI{0.0251\pm0.0041}{}\\
    & n = 16 & \SI{18.45 \pm 0.29}{} $\rightarrow$ \SI{20.53 \pm 1.21}{}
    & \SI{0.0866\pm0.0004}{} & \SI{0.0959\pm0.0012}{}
    & \SI{0.0212\pm0.0062}{}\\
    \hline
    & n = 4 & \SI{16.23 \pm 0.38}{} $\rightarrow$ \SI{17.09 \pm 1.37}{}
    & \SI{0.2200\pm0.0005}{} & \SI{0.2308\pm0.0010}{}
    & \SI{0.0310\pm0.0043}{}\\
    $\kappa = 32$& n = 8
    & \SI{17.45 \pm 0.32}{} $\rightarrow$ \SI{18.70 \pm 0.78}{}
    & \SI{0.1524\pm0.0006}{} & \SI{0.1651\pm0.0008}{}
    & \SI{0.0128\pm0.0039}{}\\
    & n = 16 & \SI{22.68 \pm 0.29}{} $\rightarrow$ \SI{25.22 \pm 1.29}{}
    & \SI{0.0866\pm0.0005}{} & \SI{0.0987\pm0.0009}{}
    & \SI{0.0206\pm0.0069}{}\\
    \hline
    & n = 4 & \SI{19.25 \pm 0.43}{} $\rightarrow$ \SI{20.78 \pm 0.85}{}
    & \SI{0.1982\pm0.0007}{} & \SI{0.237\pm0.001}{}
    & \SI{0.0129\pm0.003}{}\\
    $\kappa = \infty$& n = 8
    & \SI{19.47 \pm 0.43}{} $\rightarrow$ \SI{20.61 \pm 1.22}{}
    & \SI{0.1409\pm0.0003}{} & \SI{0.1734\pm0.0007}{}
    & \SI{0.0105\pm0.0034}{}\\
    & n = 16 & \SI{26.10 \pm 0.56}{} $\rightarrow$ \SI{29.10 \pm 1.37}{}
    & \SI{0.0795\pm0.0003}{}
    & \SI{0.1020\pm0.0008}{} & \SI{0.0113\pm0.0029}{}\\
    \hline
  \end{tabular}
  \caption{\label{table:network_parameters}%
    Measured properties of different network types. The $\zeta_\mathrm{conf}$
    value is obtained by using \Eq{zeta-recap},
    $\langle v^2 \rangle_{\text{fb}}$ is the field-based velocity squared,
    explained in \cite{axion1}, and $\langle v^2 \rangle_{\text{g}}$
    the geometrical velocity squared explained in appendix
    \ref{AppA}. The last parameter is the measured loop chopping
    coefficient $c_{\text{loop}}$.}
\end{table}

Finally, we attempt a direct determination of the loop chopping rate.
This first involves writing code to identify points along the string
and to connect them together into loops (in a periodic box all string
comes in loops).  We then identify short loops, defined as loops
whose total (invariant) length is smaller than
$\xi = \sqrt{V/\ltot}$.  We then compare to the strings output at the
previous timestep, finding the string which is closest to the loop in
question.  If the string in the previous time step is not a loop of
nearly the same size, but is instead part of a loop which stretches
well beyond the loop in question, then a loop-creation event has
occurred, and we count the length of the loop to be an amount of
cut-off loop at this time.  Note that we do \textsl{not} consider it
``loop creation'' when an existing loop of length slightly larger than
$L$ shrinks, and $L$ increases, such that the loop now qualifies as
small.  We only consider it loop chopping when a loop at time
$\tau+\Delta \tau$ corresponds to part of a larger loop at time $\tau$.
Also note that a small loop occasionally reconnects onto the network.
We do not count this as negative loop production; since it is rare and 
has almost no effect.  Finally, we average the amount of loop
production per unit volume over the last half of each evolution and
average over evolutions to estimate the statistical error bars.

\begin{figure}
  \includegraphics[width=\textwidth]{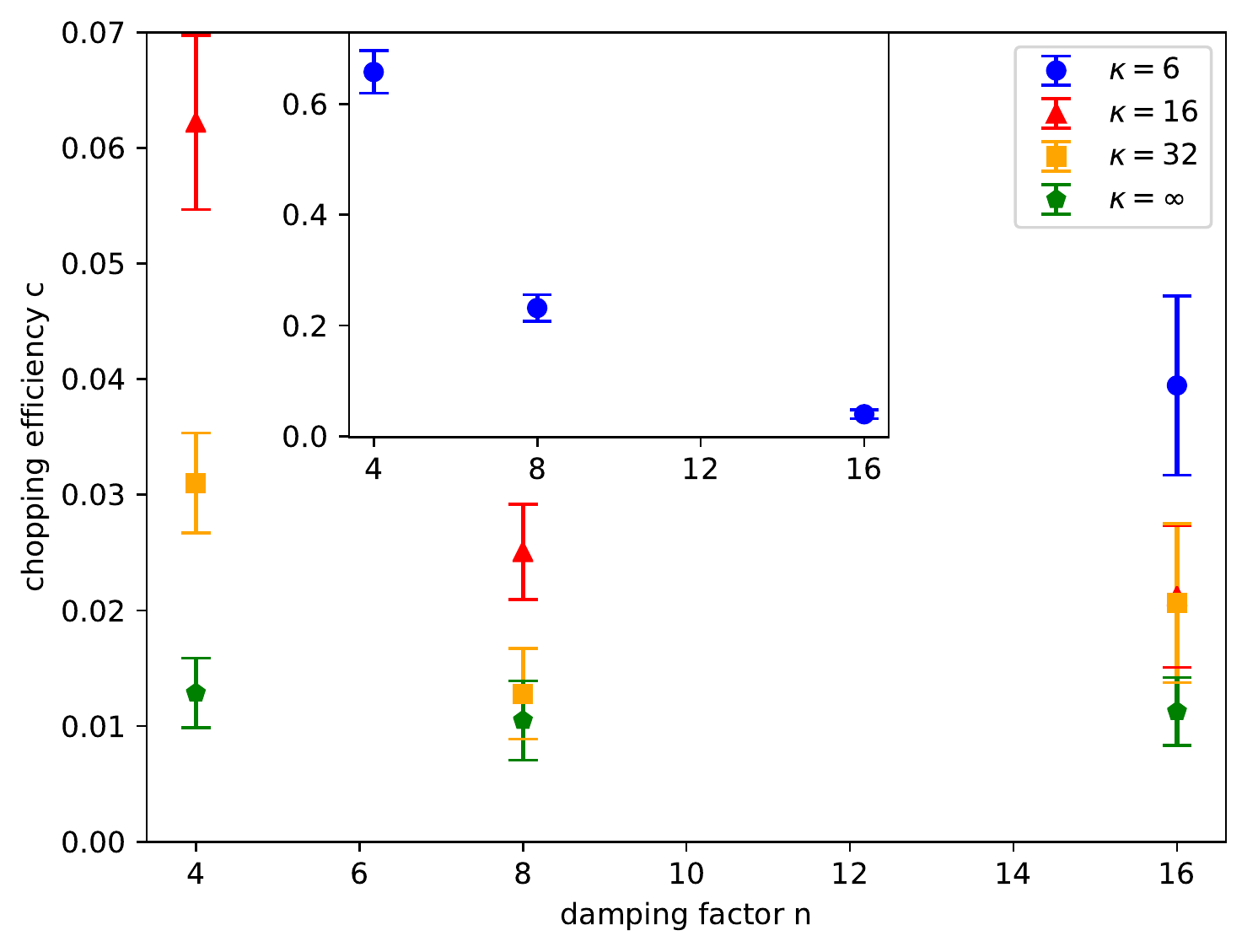}
  \caption{\label{fig:loop_chopping}
    Loop chopping efficiency for global and local string networks with various
    tensions at three Hubble-damping strengths.}
\end{figure}

The results of this study are shown in Figure
\ref{fig:loop_chopping}.  We find that loop production is more common
when the Hubble damping is weaker than when it is stronger.  We also
find that it is \textsl{much} more common for global networks,
especially those with low tension, than for local networks.  Indeed,
in our matter-dominated simulations, the chopping efficiency
$c_{\mathrm{loop}} \sim 0.6$ for scalar-only simulations, but it is
$\sim 0.013$ for abelian Higgs networks.

All of our results are summarized in Table
\ref{table:network_parameters}.

\section{Discussion:  lessons for global string models}
\label{sec:discussion}

Interactions with massless Goldstone modes play a role in the
evolution of global string networks, but are absent in local network
evolution, where only massive modes exist off the network.
We can learn more about the detailed physical impact of the
interactions with Goldstone bosons
by examining network evolution at a range of network tensions and at a
range of expansion rates (which in conformal coordinates means, a
range of Hubble damping parameters).  We have done so in this paper.

First, we showed that, within radiation-era scalar-only simulations,
there is robustly a difference between network evolution when the
``core-to-separation ratio'' is small and when it is large.  In our
language, this is comparing network evolution for radiation
domination, $n=2$, but over a range of string tensions, controlled by
the log of the core-to-separation ratio:
$\kappa = 3 \ldots 6$.  As we increase the core-to-separation ratio,
the network becomes denser and the mean string velocity becomes
smaller.  These results, especially the network density, are very
robust.

Next, by examining networks with stronger Hubble damping,
corresponding to equations of state with $w=0,-1/6,-1/4$ (or
$n=4,8,16$), we showed that this difference persists to much larger
string tensions, even when the characteristic string velocity is quite
small.  In particular, even when the damping is strong, a scalar-only
network, representing global strings with a core-to-separation ratio
of order 400, has about 1/3 the network density of a local string
network.  And the mean-squared velocity of the global network is
larger.  This is despite the fact that the mean network velocity is
very small, small enough that radiation of Goldstone modes is
predicted to play almost no role in the network dynamics.

We believe that this difference arises because inter-string forces can
play a role in accelerating the strings.  This effect does not
disappear for slowly-moving strings.  Acceleration of a string due to
Goldstone-mediated inter-string forces would add another term in
\Eq{v-evolution}, effectively modifying
$k(v) \to k(v) + g/\kappa$ with $g$ a (possibly velocity-dependent)
coefficient representing the effects of inter-string forces and
$1/\kappa$ representing that these forces become less important as the
string core's tension becomes larger.

\begin{figure}[tbh]
  \includegraphics[width=0.47\textwidth]{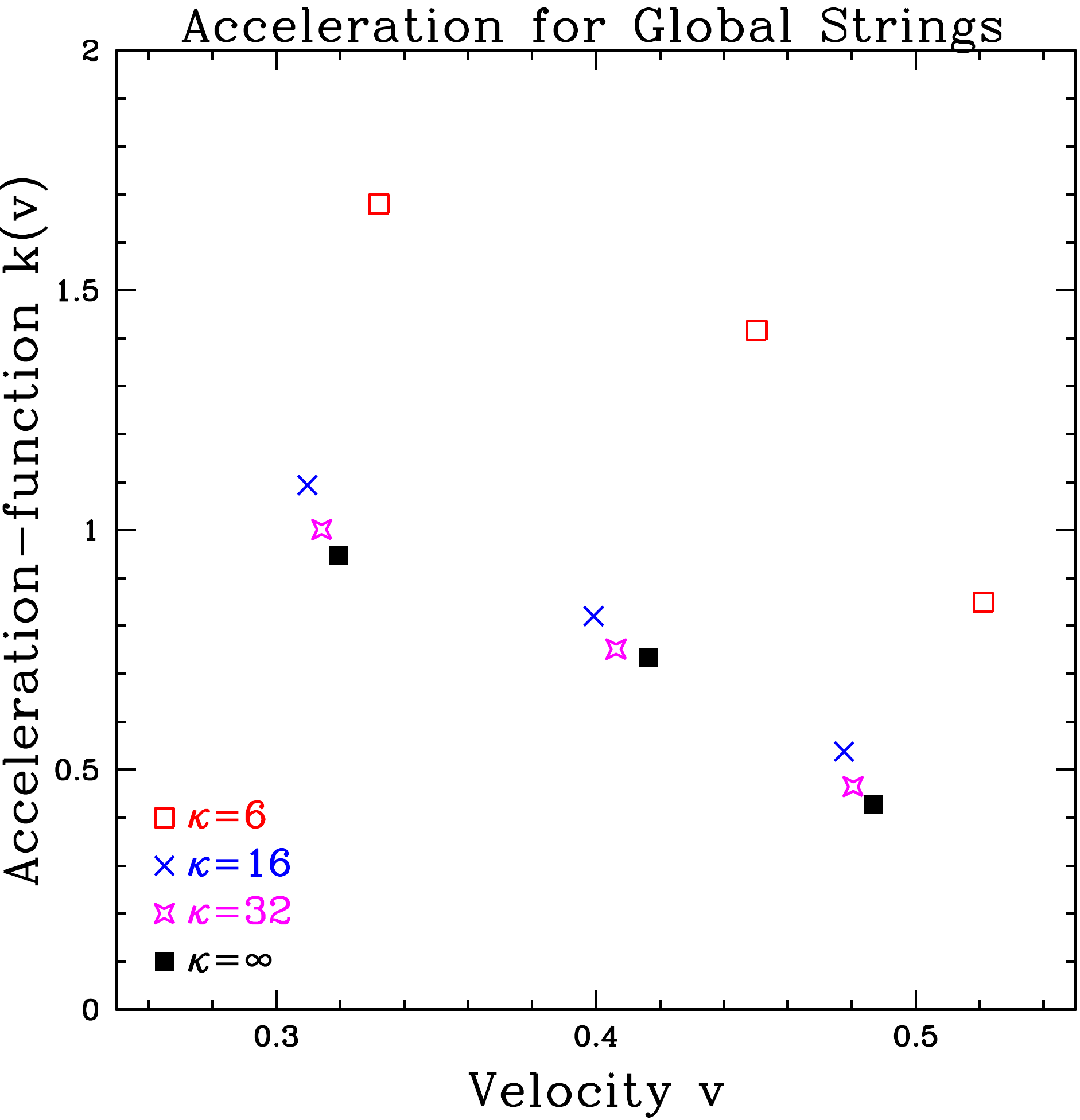}
  \hfill
  \includegraphics[width=0.47\textwidth]{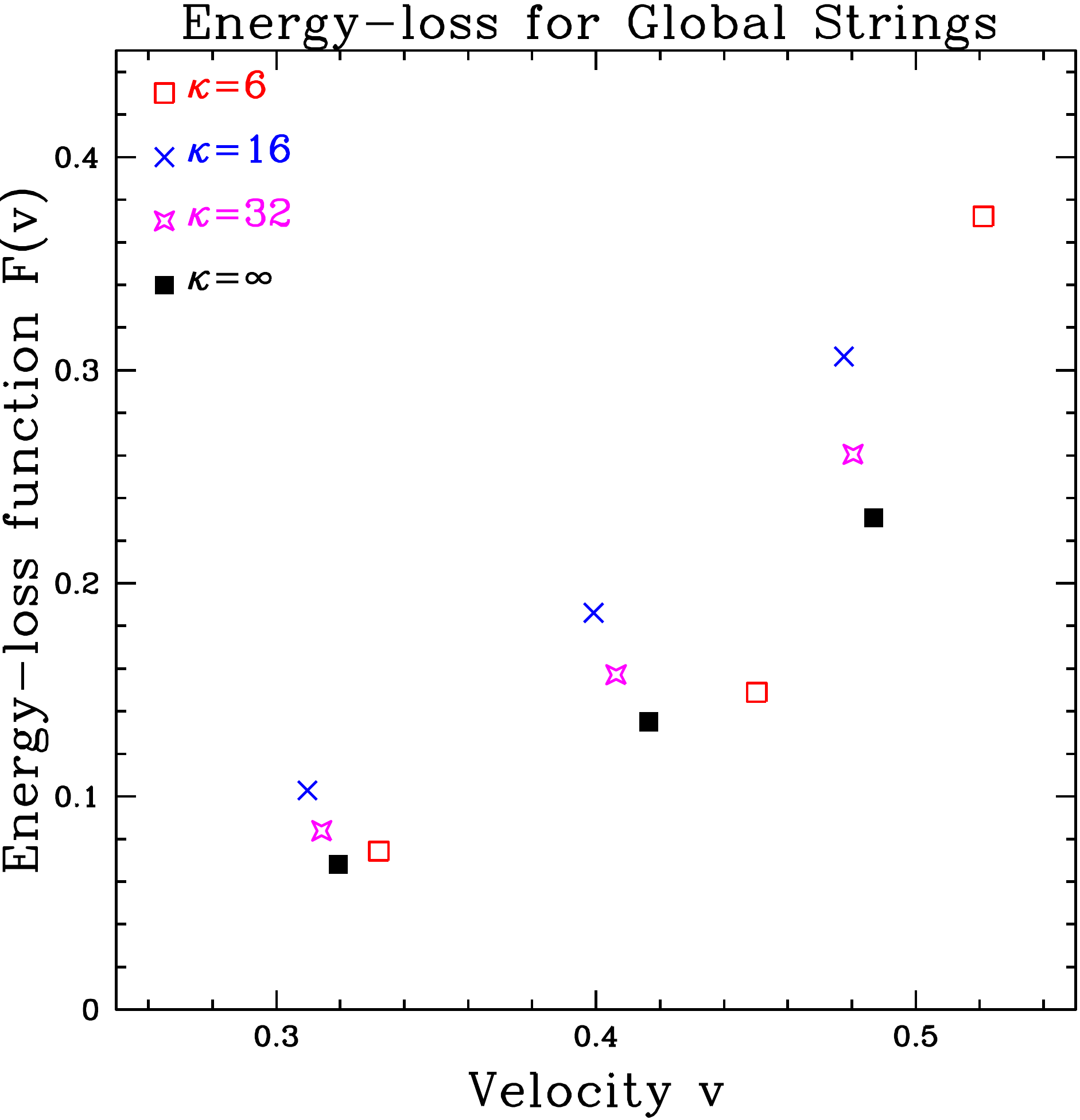}
  \caption{\label{fig:kandf}
    Observed values of the network-acceleration function $k(v)$ and
    the network energy-loss function $F(v)$, as determined using
    \Eq{dv_conformal}, \Eq{Levolve2}, and our data.  We use the
    geometrically determined network velocity.  In each figure, the
    four points at lowest/mid/highest velocity are $n=16/8/4$.}
\end{figure}

To examine this a little more, we can adopt the philosophy of
Correia and Martins \cite{Correia:2019bdl} and use our results for
$\zeta$ and $v$, together with \Eq{dv_conformal}
(rewritten as $k(v) = nv\: \xi/\tau$) and \Eq{Levolve2}, to
\textsl{determine} the ``acceleration'' function $k(v)$ and the
energy-loss or drag function $F(v)$.  Adopting our geometrically
determined network velocities $\langle v^2 \rangle_g$ from Table
\ref{table:network_parameters}, we find the results displayed in
Figure \ref{fig:kandf}.  Note that the results for $F(v)$ the drag are
rather sensitive to the network velocity.  The difference between our
geometrical and field-based velocity estimator shifts around the
points in detail within the right plot, and should be viewed as a
systematic uncertainty.  Therefore we have no convincing information
on whether $F(v)$, the efficiency with which the network radiates
energy, increases or
decreases as we pass from local to global networks.  But on the
contrary, there is a large and robust difference in the acceleration
function $k(v)$.  Because $nv\: \xi/\tau$ has only a weak dependence on
the velocity $v$, the acceleration-function $k(v)$ is rather robustly
predicted despite the differences in our string velocity estimate.
And it is dramatically different for the scalar-only simulations than
for the simulations with enhanced string tension.
A global network, especially one with a modest
separation-to-core ratio (a small $\kappa$), shows markedly more
acceleration than a global network.  This almost certainly arises
because the Goldstone modes induce inter-string forces, and it appears
to be the dominant feature leading to differences in the network
evolution, at least for the rather strongly Hubble-damped networks we
consider here.  This effect should be taken into account in future
models which seek to describe global axion networks.

\section*{Acknowledgments}

We thank the organizers of the conference ``Cosmic Topological
Defects:  Dynamics and Multi-Messenger Signatures,'' held at the
Lorentz Center in Leiden, Netherlands from 22-26 October 2018, and the
fellow participants at the meeting.  We especially thank
Carlos Martins and Jose Correia, for stimulating
discussions, as well as Mark Hindmarsh, who sent us a pre-publication
version of Ref.~\cite{Hindmarsh:2019csc} which largely stimulated
Section 3.  We also thank the GSI Helmholtzzentrum and the TU
Darmstadt and its Institut f\"ur Kernphysik for supporting this
research.

\appendix

\section{String-finding details}
\label{AppA}

Here we present a few details of how we identify the location of the
string.  Our working definition of the location of the string in pure
scalar field theory is the line of points where $\varphi$ vanishes.
The same definition works in the abelian Higgs theory.  In the
two-Higgs theory which we use to describe global strings with enhanced
tension \cite{axion3}, we can identify the string as the points where
the higher-charge field $\varphi_1$ vanishes.

This location has to be interpolated from the field values on the
actual lattice points.  We do this in three steps.  First, we identify
the plaquettes which the string penetrates.  Second, we interpolate the
$\varphi=0$ point inside each plaquette.  Third, we connect
together these plaquette-penetration points with straight-line
segments.  The remainder of this appendix will give more detail on
each of these steps.

First consider the pure scalar theory.  Label the four corners of a
plaquette $x_1=(n_x,n_y), x_2=(n_x+1,n_y), x_3=(n_x+1,n_y+1),$
and $x_4=(n_x,n_y+1)$, with scalar values
$\varphi(x_1) \ldots \varphi(x_4)$.  We define the angle between two
points as
\begin{equation}
  \theta_{i,i-1} \equiv \: \mathrm{Arg}\: \varphi(x_i) \varphi^*(x_{i-1})
\end{equation}
where Arg means, as usual, the phase in the complex plane, taken
between $-\pi$ and $\pi$.  These phases are summed as one goes around
the plaquette,
$\theta_{\mathrm{tot}} = \theta_{21} + \theta_{32} + \theta_{43} + \theta_{14}$,
and there is a string of positive/negative sense if this total is
$\pm 2\pi$, rather than zero.  Reference \cite{axion1} presents an
equivalent but more numerically efficient way to implement this
condition.

For the abelian Higgs theory, we need to make a slight modification.
To make the answer gauge invariant, we define
\begin{equation}
  \theta_{i,i-1} \equiv \: \mathrm{Arg} \left( e^{-iA}
  \varphi(x_i) \varphi^*(x_{i-1})  \right) \,,
\end{equation}
with $A$ the gauge field value%
\footnote{We use the noncompact formulation of U(1) gauge theory.}
on the link between the points $x_i,x_{i-1}$, oriented such that
$\varphi(x_i)$ and $e^{iA} \varphi(x_{i-1})$ have the same gauge
transformation properties.  The sum of the $\theta_{i,i-1}$ around the
circle now equals $-\sum A$, which is minus the magnetic field which
penetrates the plaquette, plus or minus $2\pi$ in the case that the
string penetrates the plaquette.  This algorithm originates with
Kajantie \textsl{et al} \cite{Kajantie:1998bg}.

\begin{figure}
\centerline{\includegraphics[width=0.6\textwidth]{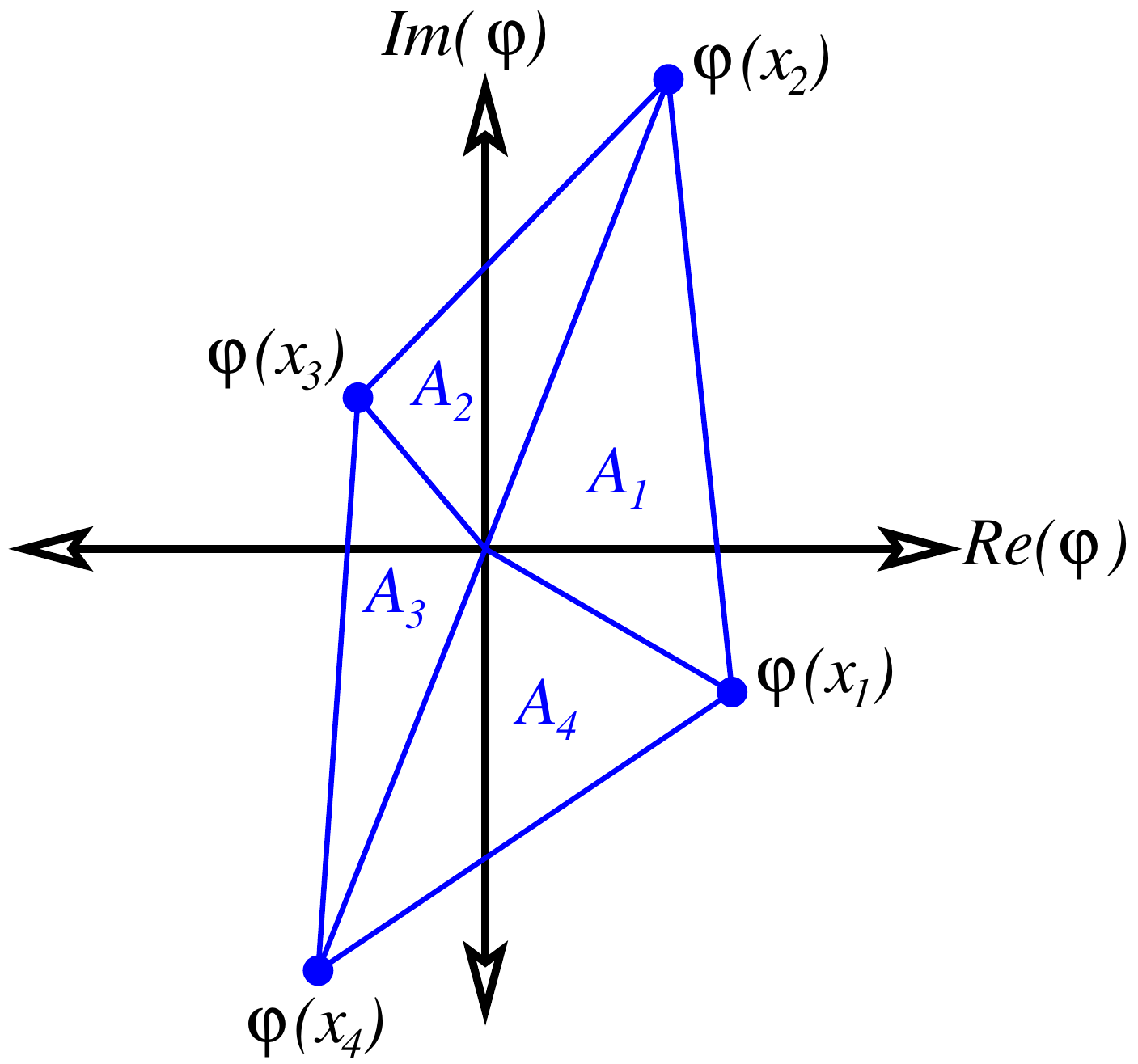}}
  \caption{\label{fig:interpolate}
    Illustration of where the four points
    $\varphi(x_1) \ldots \varphi(x_4)$ reside in the complex-$\varphi$
    plane.  The fact that the quadrangle formed by the four $\varphi$
    points encloses the origin indicates that the string penetrates
    the plaquette.  The quadrangle is not in general square, because
    the string may be moving and may not penetrate the plaquette at
    its normal.  The four areas $A_1 \ldots A_4$ can be used to
    interpolate the location of the string-crossing within the
    plaquette; the smaller $A_2$ is relative to $A_4$, the closer the
    string lies to the line (in real-space) running between $x_2$ and
    $x_3$ and the farther it is from the line (in real-space) between
    $x_4$ and $x_1$.}
\end{figure}

Having identified the plaquettes which contain a string, we must next
interpolate $\varphi(x)$ into the interior of the plaquette to find
the location where the $\varphi$ field is zero.  We start with the
pure scalar field.  Consider
the illustration in Figure \ref{fig:interpolate}, which shows where
the points $\varphi(x_1) \ldots \varphi(x_4)$ might appear as points
in the complex plane.  It is elementary to evaluate the four indicated
areas as, eg, $A_1 = \mathrm{Im}\: \varphi^*(x_1) \varphi(x_2)/2$.
We then interpolate the $x$ value of the string as
$x_{\mathrm{str}} = n_x+A_4/(A_2+A_4)$ and
$y_{\mathrm{str}} = n_y+A_1/(A_1+A_3)$.  That is, $x_{\mathrm{str}}$
gets closer to $n_x$ as the area of the triangle with corners
at $0,\varphi(n_x,n_y),\varphi(n_x,n_y+1)$ gets smaller; and it gets
closer to $n_x+1$ as the area of the triangle with corners
at $0,\varphi(n_x+1,n_y+1),\varphi(n_x+1,n_y)$ gets smaller; and
similarly in the $y$ direction.
For the case of the abelian Higgs theory or the two-scalar theory, we
use the same procedure, working in the gauge where each link $A$-field
equals $1/4$ of the magnetic flux, that is, the gauge which minimizes
the sum of squares of $A$-fields around the plaquette.

It occasionally occurs that one of the areas $A_i$ has the opposite
sign of the other three, leading to an estimated interpolated
zero-point which is outside of the plaquette.  In this case we place
the zero-point $0.001$ lattice units in from the edge of
the plaquette.

The next step is to attach these plaquette-penetrating points together
with straight line segments.  We do this as follows.  Every point
where a string goes through a plaquette carries the string from one
lattice cell (a $1\times 1\times 1$ box with lattice points as corners
and plaquettes as faces) into another; the orientation is determined
by whether the angles sum to $2\pi$ or $-2\pi$ (or whether the areas
are positive or negative).  Each cell has a total oriented number of
strings entering equal to zero.  If one string enters and one exits,
we connect the entry/exit points with a straight segment and continue
to follow the string from the exit point.  If two strings enter a cell
and two exit, we pick the pairing which leads to the shortest total
length of string within the box.

One check of our procedure is to find the total length of string,
based on the lengths of the straight segments found by our algorithm,
and to compare it to the estimate of $2/3$ of the sum of penetrated
plaquettes.%
\footnote{%
  The factor $2/3$ results by assuming that the string is locally
  straight and averaging over the directions which the string can point;
  a string pointing in the $\vec{n}$ direction penetrates
  $|n_x|+|n_y|+|n_z|$ plaquettes per unit length, which averages over
  directions to $3/2$.}
We find good agreement, whereas if we always interpolate the string to
go through the center of the plaquette we underestimate the string
length by $< 5\%$.

However, the interpolation of the location within a plaquette where
the string penetrates is not perfect, and this leads to random
fluctuations in the direction of the straight segments.  This
``renormalizes'' the dot product of the string tangent vectors by a
few percent.  In making Figure \ref{fig:angleauto}, we have applied a
multiplicative rescaling of the direction autocorrelation function
such that it approaches 1 at short distances.

\bibliographystyle{unsrt}
\bibliography{refs}

\end{document}